\documentclass{article}
\usepackage{jheppub}
\usepackage{physics}
\usepackage{tikz}

\newcommand\ads[1]{AdS$_{#1}$}
\newcommand\titlemath[1]{\texorpdfstring{#1}{Lg}}

\begin{document}

\title{The black hole/string transition in \ads{3} and confining backgrounds}

\author{Erez Y.~Urbach}
\affiliation{Department of Particle Physics and Astrophysics, Weizmann Institute of Science, Rehovot, Israel}
\emailAdd{erez.urbach@weizmann.ac.il}
\abstract{
String stars, or Horowitz-Polchinski solutions, are Euclidean string theory saddles with a normalizable condensate of thermal winding strings. String stars were suggested as a possible description of stringy (Euclidean) black holes close to the Hagedorn temperature. In this work, we continue the study initiated in \cite{Urbach:2022xzw} by investigating the thermodynamic properties of string stars in asymptotically (thermal) anti-de Sitter backgrounds. First, we discuss the case of \ads{3} with mixed RR and NS-NS fluxes (including the pure NS-NS system) and comment on a possible BTZ/string transition unique to \ads{3}.
Second, we present new ``winding-string gas'' saddles for confining holographic backgrounds such as the Witten model and determine the subleading correction to their Hagedorn temperature. We speculate a black brane/string transition in these models and argue for a possible relation to the deconfined phase of 3+1 dimensional pure Yang-Mills.
}
\maketitle

\section{Introduction}
In string theory, black holes are solutions of the low energy effective action, usually valid only for horizon size larger than the string scale $r_h \gg l_s$ (and small string coupling $g_s$). What is the correct description of a black hole microstate when it is adiabatically shrunk to the string scale?
The black hole/string transition or the correspondence principle \cite{Veneziano:1986zf,Susskind:1993ws,Horowitz:1996nw} is the idea that such stringy black holes are indistinguishable from high energy stringy bound states. In \cite{Horowitz:1997jc} a canonical ensemble version of this idea was explored (see also \cite{Damour:1999aw,Khuri:1999ez,Kutasov:2005rr,Giveon:2005jv,Chen:2021emg,Brustein:2021cza,Chen:2021dsw,Balthazar:2022szl,Bedroya:2022twb,Balthazar:2022hno}). As a thermodynamic phase of quantum gravity, the Euclidean black hole is a saddle of the string theory thermal partition on asymptotically $R^d \times S^1$.\footnote{Assuming the asymptotic circumference of the circle can be consistently frozen at the value $\beta$ \cite{Atick:1988si}.}
In terms of the temperature, this saddle is reliable for $\beta \gg l_s$. Slightly below the Hagedorn temperature of string theory $\beta_H$, 
Horowitz and Polchinski found 
\cite{Horowitz:1997jc} a new saddle, a condensate of strings winding around the thermal $S^1$ bound by gravitational interactions. We term such a saddle a ``string star'', imagining it describes a gravitational bound state of hot strings.
The black hole and the string star saddle are reliable at different regimes of temperatures, but a crude extrapolation to intermediate temperatures shows a qualitative agreement of their thermodynamic properties. The black hole/string transition can now be phrased as a precise conjecture: there exists a line of string theory backgrounds (well-defined only for small $g_s$) connecting the Euclidean black hole to the string star, and eventually (at the Hagedorn temperature $\beta=\beta_H$) to thermal $R^d\times S^1$. Technically, this line of saddles can be interpreted as a line of worldsheet CFTs connecting the Euclidean black hole to $R^d\times S^1$ through winding condensates.
In \cite{Chen:2021dsw} an argument was presented against the existence of such a line for $d>2$ type II string theory, via a worldsheet index argument. A refined version of the conjecture thus allows for the possibility of phase transitions along the line.

In \cite{Urbach:2022xzw} this discussion was extended to anti-de Sitter (AdS) spaces, by studying the properties of \ads{d+1} string stars for $d>2$. An AdS string star is a thermal winding condensate in the middle of thermal \ads{d+1}. The saddle is reliable near the AdS Hagedorn temperature $\beta_c$, $\beta-\beta_c \ll \beta_c$. As the temperature increases the solution's size grows, with a maximal size of $L \sim \sqrt{l_{ads} l_s}$. Its amplitude and free energy decrease to zero at the Hagedorn temperature, where it merges with thermal \ads{d+1}. At lower temperatures, the solution size becomes small $L\sim l_s$, at which point a correspondence principle to Euclidean AdS black holes can be drawn similar to flat space: a conjectured line of string theory backgrounds connecting the Euclidean AdS black hole solution and thermal AdS at the Hagedorn temperature via thermal winding-string condensate.\footnote{The definition of this line should be understood only in the $g_s\rightarrow0$ limit. At finite but small $g_s$ the 1-loop corrections become leading close enough to the Hagedorn temperature. As the tree level result gives an on-shell action $I \sim g_s^{-2} \left(\frac{\beta-\beta_c}{\beta_c}\right)^2$, we find the solution is trustworthy only for $(\beta-\beta_c)/\beta_c \gg g_s$.} Unlike flat space, string theory on \ads{d+1} also has a holographic dual in terms of $d$-dimensional CFT. The black hole/string transition in terms of the holographic CFT would be a conjectured line of large $N$ saddles in the thermal partition function on $S^{d-1}$, connecting the deconfined phase to the confined phase at the Hagedorn temperature (defined by the $N=\infty$ spectrum \cite{Aharony:2003sx,Harmark:2021qma}) \cite{Sundborg:1999ue,Aharony:2003sx,Alvarez-Gaume:2005dvb}.

In this work, we continue our study of AdS string stars. The first extension is the study of \ads{3} string stars ($d=2$), presented in section \ref{sec:gen_prop_ads3}. This is an interesting test for the correspondence principle that is inherently different from the flat space case. The reason is that both black holes and string stars do not exist in asymptotically flat $R^2 \times S^1_\beta$. However, in \ads{3} black hole solutions do exist and are known by the name BTZ \cite{Banados:1992wn}. Similarly, string star solutions don't exist on flat $R^2\times S^1_\beta$. But in thermal \ads{3} we find that the \ads{3} curvature stabilizes the potential for the winding condensate. As a result, thermal \ads{3} allows for reliable string star solutions close to the Hagedorn temperature.
This is an interesting check that such a correspondence is possible for a more general setting and may exist whenever black holes exist in a weakly coupled string theory.
However, contrary to higher dimensional small AdS black holes, a BTZ with small horizon radius $r_h \ll l_{ads}$ has very low temperature $\beta \gg l_{ads} \gg \beta_c$. As the BTZ gets smaller, its temperature continues to decrease further away from the Hagedorn temperature, and a simple black hole/string transition between them is impossible.

It is still possible to conjecture a different kind of correspondence.
Below temperature $\beta_c' = 4\pi^2 l_{ads}^2/\beta_c$ the BTZ solution has a tachyonic instability from strings winding the angular circle \cite{Berkooz:2007fe,Lin:2007gi,Rangamani:2007fz}. This is exactly the Hagedorn instability of thermal \ads{3}, after switching the two asymptotic circles. The natural candidate for a ``BTZ/string transition'' in \ads{3} would be conjectured line of saddles connecting the BTZ at $\beta_c'$ to thermal \ads{3} at $\beta_c$, with winding condensates at each end of the line \cite{Jafferis:2021ywg,Halder:2022ykw}. See figure \ref{fig:ads3_phase_diagram}. Notice that unlike at higher dimensions, in \ads{3} there exists a large temperature difference between the regimes of validity of the two saddles $\beta_c \ll \beta_c'$. As a result, no qualitative match can be extrapolated between the \ads{3} string star and small BTZs as a sanity check. We admit that the evidence for such a line is much weaker compared to higher dimensions.

Another interesting feature of \ads{3} string stars is the possibility in \ads{3} to turn on NS-NS flux \cite{Maldacena:1997re,Giveon:1998ns,Kutasov:1999xu,David:2002wn,Cho:2018nfn}. As the string is charged under the Kalb-Ramond field, the winding string field $\chi(x)$ is electrically charged in the effective two-dimensional description. Non-zero flux changes the solution at leading order. We study the properties of \ads{3} string stars for mixed RR and NS-NS fluxes and for the pure NS-NS system. Denoting the fraction of the NS-NS flux by $0\le \lambda \le 1$, the size of the RR flux is proportional to $\sqrt{1-\lambda^2}$. In section \ref{sec:finite_RR} we study string star solutions for finite RR flux. We find that the solutions are similar to those found in higher dimensions \cite{Urbach:2022xzw}. The NS-NS flux only contributes to the effective mass of the winding mode (at leading order in $\alpha'$), and the size of the solution is $L \sim \sqrt{l_s l_{ads}} \cdot (1-\lambda^2)^{-1/4}$. 
In section \ref{sec:small_RR} we study solutions for small RR flux $\sqrt{1-\lambda^2} \sim l_s/l_{ads}$ and for the pure NS-NS system. In this regime, the background NS-NS flux is high enough to (almost) flatten the potential for the winding mode, similar to long strings in \ads{3} with NS-NS flux \cite{Maldacena:1998uz,Seiberg:1999xz}. As a result, the size of the solution increases to the AdS scale $L \sim l_{ads}$. Moreover, 
the back-reacted NS-NS flux from the winding mode now appears at leading order, along with the metric and the dilaton.

In section \ref{sec:sub_ads} we also discuss winding string condensates in a family of confining backgrounds which includes the D3 and D4 Witten models \cite{Witten:1998qj}, the Maldacena-Nu\~{n}ez \cite{Maldacena:2000yy} and the Klebanov-Strassler \cite{Klebanov:2000hb} backgrounds. 
These gravitational backgrounds are holographically dual to theories believed to be continuously related (at low energies) to confining field theories on flat space $R^{p+1}$ ($p=2,3$). The thermodynamics of these models was studied using a semiclassical bulk analysis 
\cite{Freedman:2000xb,Gubser:2001ri,Buchel:2001gw,Buchel:2001qi,Gubser:2001eg,Buchel:2001dg,Buchel:2003ah,Aharony:2006da,Aharony:2007vg,Buchel:2007vy,Buchel:2018bzp}.
The confined phase has a ``thermal'' background dual given by analytically continuing the boundary to $R^p\times S^1_\beta$. The deconfined phase has a black-brane dual.
In this work, we find a new saddle with a thermal winding string condensate on top of the thermal background. The solution is homogeneous parallel to the spatial boundary directions $R^p$, and it is localized to the bulk IR region. We term this saddle a `winding-string gas'.
As before, it is reliable only close to the Hagedorn temperature of the holographic confining string $\beta-\beta_c \ll \beta_c$, and it is highly subdominant. We describe the properties of the solution and find the first correction to the Hagedorn temperature in these models due to the bulk curvature. It is possible to conjecture a black brane/string transition for these models, connecting the black brane saddle to the winding-string gas saddle. The phase diagram for the Witten models is similar to \ads{3}, discussed above. In this case, it is also natural to conjecture that the string winding gas is continuously related to the deconfined phase of pure $3+1$ Yang-Mills. 

Several future directions are possible. The \ads{3} calculation we performed was only at leading order in the winding condensate and at leading order in $\alpha'$. The worldsheet theory in the pure NS-NS case ($\lambda=1$) is known exactly \cite{Giveon:1998ns,Kutasov:1999xu,Maldacena:2000hw}, and it is exciting to find the string star profile exactly in $\alpha'$ \cite{Jafferis:2021ywg,Halder:2022ykw,halder_private}.
Of course, ultimately one would like to find (for the pure NS-NS background) the entire line of worldsheet CFTs interpolating between the thermal \ads{3} and BTZ. At the moment it sounds improbable, as such a deformation away from $\beta - \beta_c \ll \beta_c$ will require the entire string spectrum that couples to the winding mode, and all the orders of their interactions.
Finally, the confining backgrounds solution gives interesting predictions on its holographic dual. It would be interesting to compare these predictions to the weakly coupled regime.

\section{General properties of the \titlemath{\ads{3}} string star} \label{sec:gen_prop_ads3}
\subsection{Winding-string condensates in type II string theory} \label{sec:eft}
In this section, we will describe the low energy effective field theory (EFT) for type II string theory Euclidean saddles with a perturbative (normalizable) winding-string condensate.
This is an extension of the EFT for the winding mode together with the metric around flat space \cite{Atick:1988si,Horowitz:1997jc,Chen:2021dsw} to more general cases that include curved metric and (possibly) background fluxes such as the Kalb-Ramond flux $H_3 \ne 0$.

By ``perturbative saddle'' we mean that we start with a type II ``background saddle'' without a winding-string condensate, and will consider a small deformation of it that will include such a condensate. We will assume the undeformed metric is of the form $M_{D}\times X_{10-D}$, with $X_{10-D}$ being a $10-D$ dimensional compact space. We denote by $t\sim t+\beta$ the compact 'temporal' direction in $M_{D}$ on which we will reduce the theory, and the rest of the directions by $x^i$, $i=1,...,d=D-1$. We will assume the undeformed metric on $M_{D}$ has the form
\begin{equation}
	G = g_{tt}(x) dt^2 + g_{ij}(x) dx^i dx^j.
\end{equation}
Besides the metric, the background saddle can also include non-zero NS-NS and RR gauge fields. All the background fields are assumed to be in a gauge where they are independent of $t$, making $t$ translations a symmetry.
We set the units such that the curvature scale of the background fields is set to one $l_{curv}=1$, while keeping the $\alpha'$ explicit, and assume the background is weakly-curved $\alpha' \ll 1$.

Following \cite{Atick:1988si} we would like to consistently add an action for $\chi$, the string mode that winds once around the temporal circle $t$. We take $\chi(x)$ to be homogeneous on the compact $X_{10-D}$ directions and to depend only on $x$. Denoting the size of the normalizable $\chi(x)$ profile $L$, we will assume that it is much larger than the string scale $L \gg l_s$. Assuming the entire deformation from the saddle has $L \gg l_s$, higher order terms in the $\alpha'$ expansion of the action will be suppressed by derivatives. At leading order in $\alpha'$ the total action is
\begin{equation}\label{eq:full_action}
	I = I_\text{SUGRA} + I_\chi.
\end{equation}
The type II supergravity (SUGRA) action (at leading order in $\alpha'$) is \cite{Polchinski:1998rr}
\begin{equation}\label{eq:SUGRA_action}
	I_\text{SUGRA} = \frac{1}{16\pi G_N^{(10)}} \int d^{10} x \sqrt{G} \left[ e^{-2\Phi} \left(-R - 4\partial_\mu\Phi \partial^\mu \Phi + \frac{1}{2} |H_3|^2\right) + \frac{1}{2} |F_p|^2 \right],
\end{equation}
with $H_3 = d B_2$, $F_{p+1} = d C_p$ and $|F_p|^2 = \frac{1}{p!} F_{\mu_1 ... \mu_p} F^{\mu_1 ... \mu_p}$. $G_N^{(10)}$ is the $10$-dimensional gravitational constant. The last term stands for the possible RR $F_p$ fluxes, and we omitted the possible Chern-Simons terms for brevity.


Considering the action for the winding mode, the new ingredient compared to \cite{Urbach:2022xzw} is that here we need to describe the coupling of the winding mode to the Kalb-Ramond field. The leading interaction comes from the electric coupling of the fundamental string to the Kalb-Ramond field along the $t$ direction. It is convenient to define the 1-form gauge field 
\begin{equation}\label{eq:B_A}
 	A_i = B_{t\  i}, \text{ and } F=dA.
\end{equation} 
In these conventions for $A_i$, $\chi$ is electrically charged under the $U(1)$ winding with charge $g=\frac{\beta}{2\pi \alpha'}$.\footnote{To see this, notice the worldsheet action of the configuration $t(\sigma,\tau) = \frac{\beta}{2\pi} \sigma, x^i(\sigma,\tau)=x^i(\tau)$ is
\begin{equation}
	S_B = \frac{i \beta}{2\pi \alpha'} \int d\tau \dot x^i(\tau) B_{ti}(x^i).
\end{equation}}
Up to $O(\alpha')$ terms, the $\chi$ $d$-dimensional action takes the form
\begin{equation}\label{eq:chi_action}
\begin{split}
	I_\chi &= \frac{\beta}{16\pi G_N} \int d^d x \ \text{vol} \  e^{-2\Phi}
	\Bigg(
	|D\chi|^2 \\
	& + m^2\left(R(x)\right)|\chi|^2 + O(\alpha'^{-1}|\chi|^4)\\
	& 
	+ \left(g_1 \mathcal{R}_D[g]  + g_2 \frac{g^{tt}}{2} F_{ij} F^{ij} + \frac{g_3}{12}H_{ijk}H^{ijk}\right)|\chi|^2\Bigg) ,
\end{split}
\end{equation}
with $R(x) \equiv \frac{\beta}{2\pi} \sqrt{g_{tt}}$, $D_j = \partial_j - i g A_j$, and $g_1$, $g_2$, $g_3$ are some unknown $O(1)$ couplings. $G_N$ and 
\begin{equation}\label{eq:vol_def}
	\text{vol}=\sqrt{g}\sqrt{g_{tt}},
\end{equation}
are the $D$-dimensional gravitational constant and volume respectively.
For type II string theory the winding mode mass and the flat space Hagedorn temperature $R_H=\frac{1}{2\pi}\beta_H$ are
\begin{equation}\label{eq:m2_def}
\begin{split}
	m^2(R) = \frac{R^2-R^2_H}{\alpha'^2},\qquad &R_H^2/\alpha' = 2.
\end{split}
\end{equation}
In these notations, we also define the effective mass for the winding mode as
\begin{equation}\label{eq:m_eff}
	m^2_\text{eff}(x) = m^2(R(x)) + g^2 A_i A^i.
\end{equation}

In \eqref{eq:full_action} we added the first winding mode and the supergravity fields but neglected the rest of the higher winding modes and the rest of the massive string spectrum. This is consistent only when the mass of the winding mode is parametrically below the string scale $\alpha' m^2_\text{eff} \ll 1$ at the support of $\chi$. In practice, we need to tune the temperature of the background to be close to the Hagedorn temperature $R-R_H \ll R_H$. The second line of \eqref{eq:chi_action} includes all the $\alpha'^{-1}$ terms. After fine-tuning the quadratic mass term to be small, we still have higher order interaction terms starting at $\alpha'^{-1}|\chi|^4$. We will neglect these terms below by our assumption that the deformation is perturbative and the profile of the winding mode is parametrically small $|\chi|\ll 1$.

In the third line of \eqref{eq:chi_action} we listed the first $\alpha'^0$ terms in the action. These terms correct the mass term by a curvature scale (which also controls the Kalb-Ramond flux) amount $\Delta m^2 \sim O(1)$.\footnote{One can also write $\alpha'^0$ terms that correct the $\varphi |\chi|^2$ coupling, which are also subleading.} These terms are naively leading whenever the size of the winding mode profile is comparable to the curvature scale. Indeed some of the solutions we will find below will satisfy this condition. Nevertheless, we will argue that these terms are negligible for these cases.

Imagining a general deformation of the supergravity fields, what are the leading interaction terms? Expanding the total action around the background, the quadratic order (as in any Euclidean theory) is trivial. The leading $\alpha'^{-1}$ cubic interaction comes by expanding the kinetic term for $\chi$ to linear order around the background.
Parametrizing the deformation of the $tt$ component of the metric by $G_{tt} = g_{tt} e^{2 \varphi(x)}$, there is a universal cubic interaction by expanding the mass term
\begin{equation}
	m^2(x) |\chi|^2 = m^2(R(x)) |\chi|^2 + \frac{2R(x)^2}{\alpha'^2} \varphi |\chi|^2 + O(\varphi^2 |\chi|^2),
\end{equation}
where on the RHS $R(x)$ is of the (undeformed) background metric. Close to the Hagedorn temperature $R \sim l_s$ the interaction term is of leading $\alpha'^{-1}$ order. Whenever the background solution has a non-trivial Kalb-Ramond flux around the temporal circle $F_{ij} \ne 0$ there is also a leading cubic term from the $A_i A^i |\chi|^2$ term, of the schematic form $\frac{1}{\alpha'}\delta A_i |\chi|^2$ (we will deal with this term in the next section). However, the dilaton $\Phi$ cubic coupling to $\chi$ is of higher $\alpha'$ order due to our assumption $L \gg l_s$ and is negligible.\footnote{We assume $\Phi=0$ for the undeformed background.}
Considering a general deformation of the supergravity fields, the leading cubic interactions of $I_\text{SUGRA}$ are all with $O(1)$ (the curvature scale) couplings and can be neglected as well. Only the quadratic interactions in $I_\text{SUGRA}$ around the saddle are leading.

It is, therefore, crucial to find out which fields couple to $\varphi$ (and $\delta A_i$) at quadratic order in $I_\text{SUGRA}$. The Einstein-Hilbert term by itself gives only a kinetic term for $\varphi$, but the linear term does give a quadratic coupling to the dilaton. To see this we proceed as follows. The overall volume in \eqref{eq:SUGRA_action} is $\sqrt{G} \exp(\varphi-2\Phi)$. It is therefore useful to redefine the dilaton as 
\begin{equation}\label{eq:dilaton_def}
	\phi = \Phi-\frac{1}{2}\varphi.
\end{equation}
In this form, the quadratic order of the metric + dilaton action is given by 
\begin{equation}\label{eq:metric_dilaton_int}
\begin{split}
	\frac{1}{16\pi G_N^{(10)}} &\int d^{10} x \sqrt{G} e^{-2\Phi} \left[-R-4\partial_\mu \Phi \partial^\mu \Phi \right] \\
	&=\frac{\beta}{16\pi G_N} \int d^{d} x \text{vol} \left[ \partial_i \varphi \partial^i \varphi
	- 4\partial_i \phi \partial^i \phi 
	-2\phi \frac{\partial_i g_{tt} \partial^i \varphi}{g_{tt}} + (...)\right].	
\end{split}
\end{equation}
On the RHS we integrated over the other $10-d$ directions (including $t$) and kept only quadratic terms in $\varphi, \phi$. The last term gives a leading quadratic $\phi \partial \varphi$ interaction, that generically appears whenever $g_{tt}$ is not flat.\footnote{This is the reason this term is absent in \cite{Atick:1988si,Horowitz:1997jc,Chen:2021dsw}.} The dilaton however is expected to couple quadratically in a similar way to any supergravity field with a non-trivial profile along the $x^i$ directions. To find a saddle with a winding condensate, it is necessary to consider a general deformation of the metric and the dilaton (and other fields present in the background).

In many cases, however, there is a significant simplification that can be considered assuming the solution's length scale is also much smaller than the curvature scale $L \ll 1$. For those cases, we can further expand in orders of $L$ and neglect lower derivatives or higher powers of $x$. Looking back at \eqref{eq:metric_dilaton_int}, the kinetic terms are of order $L^{-2}$ while the interaction term is of $L^{-1}$ subleading order. In fact, in the gauge we will choose $A_i \sim O(L)$, and it is possible to show that the $\delta B |\chi|^2$ is also subleading (we will explain it for the \ads{3} case below). As a result, at leading order in $L$ the dilaton $\phi$ and the Kalb-Ramond field decouple, and we can consistently consider a perturbative deformation of the winding mode $\chi$ and the temporal metric $\varphi$ alone. The effective action in this limit will be reviewed in section \ref{sec:gen_disc}.
Whenever the size of the profile reaches the curvature scale $L\sim 1$, this approximation breaks. For those cases a general supergravity deformation is necessary. We will describe such a case in section \ref{sec:small_RR} below.

\subsection{The mixed \titlemath{\ads{3}} background} \label{sec:eom}
In this section, we begin the discussion of the \ads{3} string star.
We work in units of $l_{ads}=1$. The set of backgrounds we will consider are Euclidean 10d solutions of type IIB string theory with both NS-NS and RR fluxes, and the geometry of (thermal) $AdS_3\times S^3\times M_4$, where $M_4$ can be a four torus $T^4$ or a $K3$ manifold \cite{Maldacena:1997re,David:2002wn,Cho:2018nfn}.
The thermal \ads{3} component is given by the following background
\begin{equation}\label{eq:mixed_NS_NS_sol}
\begin{split}
	ds^2 &= d\rho^2 + \sinh^2\rho \  d\phi^2 + \cosh^2\rho \ dt^2,\\
	B_2 &= i \lambda \sinh^2\rho \  d\phi \wedge dt,\\
	C_2 &= i \sqrt{1-\lambda^2} \sinh^2\rho \  d\phi \wedge dt,
\end{split}
\end{equation}
with $t\sim t+\beta$ and $0\le \lambda\le 1$.\footnote{The quantization of the fluxes (on $S^3$) requires both $\lambda / \alpha'$, $\sqrt{1-\lambda^2} / \alpha'$ to be integers. In this work, we expand at leading order in $\alpha'$, and therefore this quantization is invisible \cite{Maldacena:1997re,Cho:2018nfn}.} The gauge fields $B,C$ also have fluxes from the $S^3$ component, which we omitted.
$\lambda = 0$ is the pure RR solution, for which string stars solutions were studied in \cite{Urbach:2022xzw}, and $\lambda=1$ is the pure NS-NS solution. We note that the latter can be studied not only for type IIB but in any closed string theory, and it is also known exactly in $\alpha'$ via an $\text{SL}(2,R)$ WZW worldsheet description \cite{Giveon:1998ns,Kutasov:1999xu,Maldacena:2000hw}. As in \cite{Urbach:2022xzw} $\beta$ is also the holographic temperature in the dual CFT on $S^1_{2\pi}\times S^1_\beta$.

In these conventions the background gauge field \eqref{eq:B_A} is $A_\phi = i \lambda \sinh^2\rho$ and $R(\rho) = R_0 \cosh(\rho)$, with $R_0 = \frac{\beta}{2\pi}$. The (background) effective mass \eqref{eq:m_eff} is given by
\begin{equation}\label{eq:ads3_m2_eff}
	m^2_\text{eff}(\rho) = \frac{1}{\alpha'^2} \left(R_0^2 \left(\cosh^2\rho - \lambda^2 \sinh^2\rho \right) -R_H^2\right).
\end{equation}
For self consistency we assumed scale separation $\alpha' m^2_\text{eff}(\rho) \ll 1$, which in this case gives $R_0-R_H \ll R_H$ for $\rho \sim O(1)$. The EFT is therefore applicable only for temperatures close to the Hagedorn temperature.

Following the previous section, to write the two-dimensional EFT we need to expand $I_\text{SUGRA}$ to quadratic order around a general supergravity deformation.
Assuming the spherical symmetry of the solution, we are instructed to consider a deformation for every field with a nontrivial background profile along the $\rho$ direction.
A general spherically symmetric deformation of the \ads{3} components of \eqref{eq:mixed_NS_NS_sol} has the form\footnote{We gauge fixed the field potential $B_2$, $C_2$ to also depend solely on $\rho$, which fixes the allowed components.}
\begin{equation}\label{eq:deform}
\begin{split}
	\Phi &= \phi + \frac{1}{2} (\varphi + \psi),\\
	G &= d\rho^2+e^{2\varphi} \cosh^2\rho \ dt^2 + e^{2 \psi} \sinh^2\rho \  d\phi^2,\\
	B_2 &= i \lambda e^{\zeta} \sinh^2\rho \  dt \wedge d\phi,\\
	C_2 &= i \sqrt{1-\lambda^2} e^{\eta} \sinh^2\rho \  dt \wedge d\phi,
\end{split}
\end{equation}
with $\phi, \varphi, \zeta$ and $\eta$ all function of $\rho$. In particular, we don't need to consider deformations of other RR fluxes. Similar to \eqref{eq:dilaton_def}, we redefined $\Phi$ in such a way to satisfy
\begin{equation}
	\sqrt{G} e^{-2\Phi} = \text{vol} \  e^{-2\phi},
\end{equation}
with $\text{vol}=\cosh \rho \ \sinh \rho$ the volume form of the background metric. 

Following the discussion in the previous section, we assume small back-reaction $|\phi|,|\varphi|,|\psi|,|\zeta|, |\eta|\ll 1$ and consider only its leading coupling to $\chi$.
Expanding the winding mode action \eqref{eq:chi_action} to ($\alpha'^{-1}$) cubic order, and integrating over the angular direction, gives
\begin{equation}\label{eq:chi_action_ads3}
\begin{split}
	I_\chi &= \frac{2\pi \beta}{16\pi G_N}\int d\rho \ \text{vol} \left[ 
		|\chi'|^2+m^2_\text{eff}(\rho) |\chi|^2 + \frac{2 R_0^2}{\alpha'^2} \left( 
		\cosh^2\rho \  \varphi -\sinh^2\rho \  (\zeta-\psi)\right) |\chi|^2
	\right],
\end{split}
\end{equation}
with $G_N$ being the 3-dimensional gravitational constant, and the effective mass \eqref{eq:ads3_m2_eff}. Here and below, primes are derivatives with respect to $\rho$. In particular, we find both $\varphi |\chi|^2$ and $(\zeta-\psi) |\chi|^2$ terms of the same leading $\alpha'^{-1}$ order. 
Substituting the deformation \eqref{eq:deform} in the SUGRA action \eqref{eq:SUGRA_action} gives to quadratic order
\begin{equation}\label{eq:SUGRA_action_ads3}
\begin{split}
	I_\text{SUGRA}^{(2)} &= \frac{2\pi \beta}{16\pi G_N^{(3)}} \int d\rho \ \text{vol} \  \Bigg[
	-4 (\phi')^2 +(\varphi')^2 +(\psi')^2 \\
	&-\lambda^2\left(\frac{1}{2}\tanh^2\rho \ (\zeta')^2+4(\psi+\varphi-\zeta)^2-4\tanh\rho \ \zeta'(\psi+\varphi-\zeta)\right)\\
	&-(1-\lambda^2)\left(\frac{1}{2}\tanh^2\rho \ (\eta')^2
	+ 4\eta^2-4\eta (\varphi+\psi)-2\tanh\rho \ \eta'(\psi+\varphi-2\eta)
	\right)\\
	&-4\phi\left(\tanh\rho \ \varphi'+\coth\rho \ \psi'
	+\lambda^2 (2\varphi+2\psi-2\zeta-\tanh\rho \ \zeta')
	\right)
	\Bigg].
\end{split}
\end{equation}
To avoid a curvature singularity at $\rho=0$ we need to maintain $\psi(0)=0$. Regularity of the fields at $\rho=0$ also sets $\chi'(0)=\phi'(0)=\varphi'(0)=\psi'(0)=\zeta'(0)=\eta'(0)=0$. At $\rho=\infty$ all the fields should vanish for normalizability.

\subsection{Thermodynamics and the BTZ/string transition} \label{sec:thermo}
The total EFT action compared to that of thermal \ads{3} is the sum of \eqref{eq:chi_action_ads3} and \eqref{eq:SUGRA_action_ads3}
\begin{equation}\label{eq:ads3_action}
	I_\text{String star} -I_{\text{TAdS}_3} = I^{(2)}_\text{SUGRA}+I_\chi.
\end{equation}
The \ads{3} string star is the bound state saddle of \eqref{eq:ads3_action}. Being a solution of the action, it is a Euclidean saddle of the (Euclidean) string theory partition function with asymptotic thermal \ads{3} boundary condition.
As a Euclidean saddle, it has free energy of order $G_N^{-1}$, just like black holes. It is applicable only near the Hagedorn temperature $R_0-R_H \ll R_H$.

In $d=2$ the AdS black hole solution is known by the BTZ solution \cite{Banados:1992wn}. It's on-shell action, compared to the thermal \ads{3}, is (in units $l_{ads}=1$)
\begin{equation}
	I_\text{BTZ} -I_{\text{TAdS}_3} = -\frac{\pi}{4 G_N} \left(R_0^{-1}-R_0\right).
\end{equation}
Thus, thermal \ads{3} is canonically dominant for low temperatures $R_0 > 1$, while the BTZ is dominant for high temperatures $R_0 < 1$.
Whenever the string star saddle is reliable $R \sim R_H \ll 1$ the dominant saddle is the BTZ. The string star saddle is therefore non-perturbatively unstable, and exponentially subdominant canonically. As we will see numerically in figure \ref{fig:free_energy} below (and also in \cite{Urbach:2022xzw}) the string star free energy \eqref{eq:ads3_action} is positive (note that the expression is not clearly positive due to the $\zeta, \eta$ kinetic terms \eqref{eq:SUGRA_action_ads3}). Therefore it is subdominant also compared to the thermal \ads{3} solution.

For higher dimensional \ads{d+1} $d>2$ an AdS black hole/string transition was suggested between the AdS string star saddle and AdS black holes \cite{Urbach:2022xzw}. The argument was carried from the black hole/string transition previously suggested between the flat-space string star saddle (or the Horowitz-Polchinski solution) and the (Euclidean) Schwarzschild saddle \cite{Horowitz:1997jc,Chen:2021dsw}. The reason is that at the two ends of the correspondence point around $R_0 \sim l_s$, both the AdS black hole and the AdS string star are of size much smaller than the AdS scale and can be well approximated by their asymptotically flat counterparts.
\begin{figure}[t]
	\centering
	\includegraphics[width=.9\linewidth]{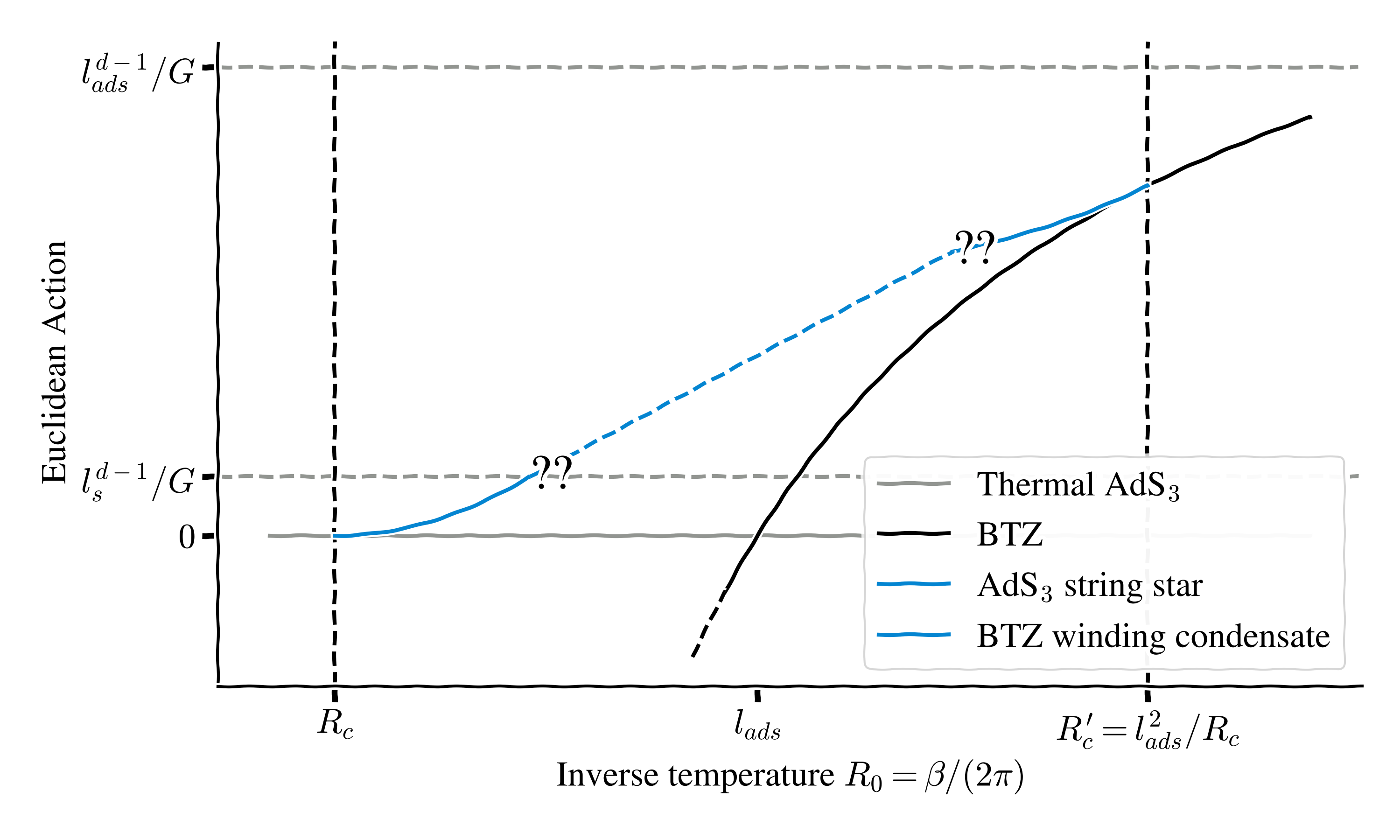}
	\caption{A conjectured phase diagram for \ads{3} (for $l_{ads} \gg l_s$). The Euclidean action as a function of the inverse temperature $R_0$ is sketched. The gray and black lines describe the thermal \ads{3} and the BTZ respectively. The string star saddle is the blue line close to the Hagedorn temperature $R_c$. Its dual, a BTZ winding condensate, is the blue line connected to the BTZ saddle at $R_c'$. A BTZ/string transition in \ads{3} would be a hypothetical line of saddles connecting the two (dashed blue line).
	We note that the sketch should be understood in the limit $g_s \rightarrow 0$. At finite but small $g_s$ the string star saddle can't be trusted arbitrarily close to $R_c$ whenever the 1-loop effect is leading.}
	\label{fig:ads3_phase_diagram}
\end{figure}

In $d=2$ however asymptotically flat Schwarzschild solutions simply don't exist, so a similar argument can't be drawn. Going back to \ads{d+1},  AdS black holes exist only for $R_0 < l_{ads}/\sqrt{d(d-2)}$. For small enough horizon radius $r_h / l_{ads} < \sqrt{\frac{d-2}{d}}$ the temperature grows as the radius shrinks, just like in flat space. 
This is not the case for $d=2$. The temperature of the BTZ solution always grows with the radius $R_0 = 1 / r_h$.
As a result, any attempt to connect the \ads{3} string star to a small black hole close to the Hagedorn temperature is impossible, because at high temperatures only large black hole solutions exist.
A clue comes when we consider the asymptotic symmetries of the question. Any worldsheet with asymptotic $S^1_\phi \times S^1_\beta$ has two $U(1)$ windings. The thermal \ads{3} saddle breaks the $S^1_\phi$ winding spontaneously, while the BTZ saddle breaks the $S^1_\beta$ winding. The \ads{3} string star saddle on the other hand (spontaneously) breaks both winding symmetries. To connect the \ads{3} string star saddle to the BTZ we need to consider deformations of the BTZ by $S^1_\phi$ winding strings.

Denoting the \ads{3} Hagedorn temperature by $R_c$, the BTZ solution has a known instability at $R_c' = 1/R_c$ \cite{Berkooz:2007fe,Lin:2007gi,Rangamani:2007fz}. This instability is exactly the tachyon instability of the thermal \ads{3} at the AdS Hagedorn temperature after we switch the roles of the two asymptotic circles ($R_0 \mapsto 1/R_0$). In other words, below $R_0>R_c'$ strings winding around the $\phi$-circle are tachyonic. Considering the \ads{3} string star saddle after switching the circles we get a new saddle that describes a normalizable $\phi$-winding condensate in the BTZ background. This saddle also breaks both winding symmetries spontaneously, but it is connected (in its regime of validity $R_c'-R_0 \ll R_c'$) to the BTZ instead of the thermal \ads{3}. In terms of global symmetries pattern, it is therefore natural to speculate these two branches meet around $R_0 \sim 1$.\footnote{By `meet' we mean through a line of string theory backgrounds labeled by the asymptotic temperature $\beta = 2\pi R_0$. } 
This phase diagram for \ads{3} was also suggested in \cite{Halder:2022ykw}. A schematic plot of this suggested picture is drawn in figure \ref{fig:ads3_phase_diagram}.

\section{Sub-AdS solutions and confining strings} \label{sec:sub_ads}
\subsection{General discussion} \label{sec:gen_disc}
This section studies general properties of holographic thermal confining theories, in cases where the ``IR wall'' is geometrically realized by a contractible (non-thermal) cycle. Thermal $AdS_{d+1}$ is one such example \cite{Urbach:2022xzw}, in which an $S^{d-1}$ closes in the IR region. We start by showing that around such a background there exists a saddle of thermal winding string condensate that depends (to leading order) only on the curvature scale at the IR point at which the cycle closes.
The reason would be that for this background, the size of the solution would be much smaller than the characteristic curvature scale. In terms of AdS$_{d+1}$ these solutions satisfy $L \ll l_{ads}$ and were reviewed in \cite{Urbach:2022xzw}. The analysis allows us to find the characteristic behavior of the solution and the leading correction to the Hagedorn temperature in the dual holographic theory.

The topology of the geometries we will study is $S^1_\beta \times \mathbb{R}^{d} \times \mathcal{M}$. $S^1_\beta$ asymptotes to the thermal cycle of the holographic theory, $\mathbb{R}^{d}$ is made of a radial direction $r$ and a shrinking $S^{d-1}$, and $\mathcal{M}$ is some (perhaps non-compact) manifold. We further assume the solution preserves $SO(2) \times SO(d)$ symmetry. The string metric close to the shrinking point $r=0$ of the $S^{d-1}$ sphere is, to subleading order in $r^2$,
\begin{equation} \label{eq:conf_ansatz}
	ds^2 = dr^2 + \left( 1 + \frac{r^2}{l^2}\right) dt^2 + r^2 d\Omega^2_{d-1} + ds^2_{\mathcal{M}}
\end{equation}
The Euclidean time direction satisfies $t \sim t +2\pi R_0$. Higher orders of the metric are assumed to be down by higher orders of $r^2/l^2$ with $l$ being the curvature scale of the (IR) geometry. As a result, we implicitly assume $r \ll l$ and $l_s \ll l$. On general grounds the parameter $R_0$ is related linearly to the holographic theory temperature $\beta$. We also assume that for $r \ll l$ the internal space geometry $ds^2_\mathcal{M}$ is independent of $r$.
In this geometry, and for temperatures close to the Hagedorn temperature, we look for non-trivial winding solutions that are homogeneous both on $\mathcal{M}$ and on the closing $S^{d-1}$ sphere. In cases that $\mathcal{M}$ is non-compact these solutions are dual (in the holographic theory) to a gas of interacting (confining) strings, and so we term them ``winding-string gas'' solutions.

In section \ref{sec:eft} we explained that whenever the winding mode profile size is parametrically smaller than the curvature scale $L \ll l$ we can expand the EFT in orders of $L/l$ and consider only the winding mode $\chi$ and the temporal metric deformation $\varphi$. In practice, we take only the highest number of derivatives (within the $O(\alpha')$ action) and fewer powers of $r$. Assuming both $\chi(r),\varphi(r)$ depend only on $r$, the leading $O(l^2/L^2)$ terms in \eqref{eq:full_action} are
\begin{equation} \label{eq:small_sol_eft}
	I_{L \ll l} = \frac{\beta V_{d-1}}{16\pi G_N} \int r^{d-1} dr \left[(\varphi')^2 + |\chi'|^2 + m^2_\text{eff}(r) |\chi|^2 + \frac{2R_0^2}{\alpha'^2} \varphi |\chi|^2 \right],
\end{equation}
with $V_{d-1}$ the volume of a unit $S^{d-1}$ sphere.
At leading order for small $r$ it is enough to expand the effective mass to quadratic order in $r$. Assuming no Kalb-Ramond flux in the $t$ direction the effective mass \eqref{eq:m_eff} is
\begin{equation}\label{eq:conf_m_eff}
	m^2_\text{eff} (r) = \frac{1}{\alpha'^2}\left(R_0^2 - R_H^2 + \frac{R_0^2 r^2}{l^2} \right).
\end{equation}
Notice that the only $l$ dependence in the action \eqref{eq:small_sol_eft} is through \eqref{eq:conf_m_eff}. Whenever a (non-singular) Kalb-Ramond flux exists it only corrects the $r^2$ term and can be swallowed into a redefinition of $l$. Naively the second term is subleading for $L\ll l$, but for temperatures close enough to the Hagedorn temperature $(R_0^2-R_H^2)/R_0^2 \sim L^2/l^2$ it is of the same order and should be added (while higher orders in $r$ are still subleading). The resulting equations for $\chi(r), \varphi(r)$ are 
\begin{equation}\label{eq:conf_eom}
\begin{split}
	\chi ''+ \frac{d-1}{r} \, \chi ' - \left(m^2_\text{eff}(r)+ \frac{4}{\alpha'} \varphi \right) \chi & = 0, \\
	\varphi ''+ \frac{d-1}{r} \, \varphi ' - \frac{2}{\alpha'} |\chi|^2 & = 0.
\end{split}
\end{equation}
These equations are of the same form found in the near-Hagedorn limit at \cite{Urbach:2022xzw}.

The winding EFT regime of validity for the winding EFT is $(R_0^2-R_H^2)/R_H^2 \ll 1$. In this section, we further restrict ourselves to $(R_0^2-R_H^2)/R_H^2 \sim l_s/l$. We will denote this regime of temperatures as ``near-Hagedorn'' and parameterize it explicitly by defining
\begin{equation}
	R_0^2 \equiv R_H^2 + \delta \cdot \frac{l_s^{3}}{l}.
\end{equation}
We can further reparameterize 
\begin{equation}
	\chi(r) = \frac{l_s}{l} \cdot \hat \chi(r/L), \quad \varphi(r) = \frac{l_s}{l} \cdot \ \hat \varphi(r/L),
\end{equation}
 with $L^2\equiv l_s l$, to get ($x=r/L$)
\begin{equation}\label{eq:conf_full_eom}
\begin{split}
	\hat \chi ''+ \frac{d-1}{x} \, \hat \chi ' - \left(\delta + 2 x^2 + 4 \hat \varphi \right) \hat \chi & = 0, \\
	\hat \varphi ''+ \frac{d-1}{x} \, \hat \varphi ' - 2 |\hat \chi|^2 & = 0,
\end{split}
\end{equation}
with $'$ denote an $x$ derivative.
These equations were studied numerically in \cite{Urbach:2022xzw}.
Because \eqref{eq:conf_full_eom} is dimensionless we find $|\chi|,|\varphi| \sim l_s/l$ and solution's size $L \sim \sqrt{l_s l}$. Because (for the original metric to be weakly curved) $l_s/l \ll 1$ the solutions for \eqref{eq:conf_full_eom} are all self-consistent. Specifically note that $l_s \ll L\ll l$, justifying the approximation \eqref{eq:small_sol_eft}.
Numerically it was found that the solution's size increases with the temperature \cite{Urbach:2022xzw}. 

The Hagedorn temperature of the theory is defined as the temperature where the linearized equations have a normalizable solution, giving rise to a normalizable zero mode in the partition function. At the order we work with, this is given by $\delta_c = -\sqrt{2} d$, and the corresponding normalizable mode is $\chi(r) = \exp(-r^2/(\sqrt{2} L^2))$. In other words, the leading curvature correction to the Hagedorn value for $R_0$ is
\begin{equation}\label{eq:R_c}
	R_c^2  = 2 \alpha' \left(1 - \frac{d}{\sqrt{2}} \frac{l_s}{l} + O(l_s^2/l^2)\right).
\end{equation}

However, $R_0$ is not directly related to the holographic theory temperature $\beta$. In the examples below the geometry is given by an analytic continuation from a Lorentzian manifold with a boundary of $R^{p,1}$ for some $p$. The Lorentzian solutions take the general form
\begin{equation}
	ds^2 =  dr^2 + f(r)\left( -(dx^0)^2 + \sum_{i=1}^p (dx^i)^2\right) + (...),
\end{equation}
where $r$ is again the radial bulk direction, $x^0,...,x^p$ are parallel to the boundary $R^{p,1}$, and in the bracket we ignored the rest of the coordinates. The Euclidean solution is then taken by analytically continuing $x_0 = i t$, with the same $t \sim t+\beta$ as in \eqref{eq:conf_ansatz}. Comparing to \eqref{eq:conf_ansatz} gives $f(r=0) = 4\pi^2 R_0^2 / \beta^2$ at $r=0$. The confining string tension $T_{st}$ is given by calculating the bulk worldsheet action in the $t, x^i$ IR plane ($r=0$) measured in units of the holographic boundary lengths given by $\beta$:
\begin{equation}
	T_{st} = \frac{1}{2\pi \alpha'}\sqrt{g_{tt} g_{xx}}\mid_{r=0} = \frac{f(0)}{2\pi \alpha'}=\frac{2\pi R_0^2}{\alpha' \beta^2}.
\end{equation}
Together with \eqref{eq:R_c}, the leading curvature correction to the holographic Hagedorn temperature is
\begin{equation} \label{eq:conf_hag}
	\beta_c = \sqrt{\frac{4\pi}{T_{st}}} \left(1-\frac{d}{2\sqrt{2}} \frac{l_s}{l} + O(l_s^2/l^2) \right).
\end{equation}

\subsection{\titlemath{\ads{3}} with finite RR flux} \label{sec:finite_RR}
We now go back to study the \ads{3} string star solutions in the background \eqref{eq:mixed_NS_NS_sol}. Following the previous section we will when small solutions $L \ll l_{ads}$ exist. Assuming $\rho \ll l_{ads}$ we can directly expand \eqref{eq:SUGRA_action_ads3} and \eqref{eq:chi_action_ads3} to $L^{-2}$ order and see that all the fields besides $\varphi$ decouple, leaving us with \eqref{eq:small_sol_eft} (at $d=2$). In particular, notice that while both $\varphi$ and $\zeta$ couple to $\chi$ at the same $\alpha'^{-1}$ order \eqref{eq:chi_action_ads3}, the $\zeta$ coupling is smaller by a factor of $\rho^2/l_{ads}^2 \ll 1$.
Expanding the effective mass \eqref{eq:ads3_m2_eff} to leading order gives (bringing back the $l_{ads}$ units)\footnote{The discussion here is for $\lambda \sim O(1)$. We note that for small enough $\lambda \sim l_s/l_{ads}$ the $\lambda$ (and the entire NS-NS flux) dependence is absent at leading order in $\alpha'$, and appears together with the suppressed higher $\alpha'$ terms in \eqref{eq:chi_action}.} 
\begin{equation}
	m^2_\text{eff}(\rho) = \frac{1}{\alpha'^2}\left(R_0^2 + \frac{(1-\lambda^2) R_0^2}{l_{ads}^2}\rho^2 - R_H^2\right).
\end{equation}
This is the same form as in \eqref{eq:conf_m_eff} only with the effective scale $l = l_{ads}/\sqrt{1-\lambda^2}$. We can follow the same steps as in the previous section and assume a small solution $l_s \ll L \ll l$. 
We also note that as the winding mass of order $m^2 \sim l_{ads}/l_s^3$, we can consistently neglect higher-curvature contributions to the mass which are of subleading order $\Delta m \sim l_{ads}^{-2}$.

Following the previous section, for near-Hagedorn temperatures $(R_0^2-R_H^2)/R_H^2 \sim l_s/l_{ads}$ the \ads{3} string star solution has an amplitude of
\begin{equation}
	|\chi|,|\varphi| \sim \frac{l_s}{l_{ads}} \sqrt{1-\lambda^2},
\end{equation}
and a characteristic size
\begin{equation}
	L \sim \sqrt{l_s l_{ads}} (1-\lambda^2)^{-1/4}.
\end{equation}
In this geometry $\beta = 2\pi R_0$ and therefore, following \eqref{eq:conf_hag}, the leading correction to the \ads{3} Hagedorn temperature is
\begin{equation} \label{eq:finite_RR_R_c}
	R_c^2 = 2 \alpha' - 2^{3/2} \sqrt{1-\lambda^2} \frac{l_s^3}{l_{ads}} + O(\alpha'^2/l_{ads}^2).
\end{equation}
In \cite{Urbach:2022xzw} the leading correction for the Hagedorn temperature in $AdS_5\times S^5$ was matched with an integrability analysis of $\mathcal{N}=4$ SYM \cite{Harmark:2021qma}.
It would be interesting to find a way to similarly compare \eqref{eq:finite_RR_R_c} to the recent integrability results for the pure RR \ads{3} case ($\lambda=0$) \cite{Cavaglia:2021eqr,Ekhammar:2021pys,Cavaglia:2022xld,Frolov:2021bwp}.

The thermodynamic stability of these solutions was studied in \cite{Urbach:2022xzw}. $\zeta$ is negligible in \eqref{fig:free_energy} and the string star free energy can be shown to be positive. This saddle is therefore non-perturbatively unstable compared to the thermal \ads{3} saddle (and to the BTZ saddle).

This discussion is self-consistent only when the size of the solution is smaller than the AdS curvature scale, $L \ll l_{ads}$, or $1-\lambda^2 \gg \alpha'/l_{ads}^2$.
Note that $\sqrt{1-\lambda^2}$ is the magnitude of the RR flux $F_3$ \eqref{eq:mixed_NS_NS_sol}.
We, therefore, conclude that for near-Hagedorn temperatures $(R_0^2-R_H^2)/R_H^2 \sim l_s/l_{ads}$ and large enough RR-flux $1-\lambda^2 \gg \alpha'/l_{ads}^2$ the leading behavior of the mass around $\rho=0$ solely controls the profile of the string star. As we lower the RR flux the profile size is larger and the amplitude is smaller. Around $1-\lambda^2 \sim \alpha'/l_{ads}^2$ the solution size reaches the AdS scale and the approximation we took breaks. We will discuss this case in section \ref{sec:small_RR} below.

For lower temperatures, the solution size turns shorter. 
Below the near-Hagedorn regime, but for temperatures in which the EFT is still valid $l_s/l_{ads} \ll (R_0^2 - R_H^2)/R_H^2 \ll 1$ one might expect reliable solutions that are arbitrarily short (although with $L \gg l_s$). This is not the case and for a good reason. If arbitrarily small solutions would exist, they would approximately solve the flat space equations for $d=2$, and it is known that no such solutions exist \cite{Horowitz:1997jc,Chen:2021dsw}. Instead, whenever the solution is reliable ($|\chi|,|\varphi|\ll 1$) its size remains comparable to $L \sim \sqrt{l_s l_{ads}}$.
In other words, this section describes the behavior of \ads{3} string stars whenever they exist for finite RR flux $1-\lambda^2 \gg \alpha'/l_{ads}^2$.

\subsection{The D4 Witten model} \label{sec:d4_back}
Consider $4+1$ dimensional maximally symmetric $SU(N)$ Yang-Mills theory (with coupling $g_5$) compactified on a circle of radius $R$ with anti-periodic boundary condition for the fermions. We denote the $4+1$ ’t~Hooft coupling by $\lambda_5 = g_5^2 N$.
It is holographically dual to the near horizon limit of $N$ D4 branes \cite{Witten:1998zw}, with the $9+1$ (type IIA) metric
\begin{equation}\label{eq:witten_d4}
\begin{split}
	ds^2 & = \left(\frac{u}{R_{D4}}\right)^{3/2} \left[ -(dx^0)^2 + \sum_{i=1}^3 (dx^i)^2 +f(u) (dx^4)^2 \right] + \left(\frac{R_{D4}}{u}\right)^{3/2} \left[\frac{du^2}{f(u)} + u^2 d\Omega_4^2 \right]\\
	F_{(4)} &= \frac{2\pi N}{V_4} \epsilon_4, \quad e^\phi = g_s \left(\frac{u}{R_{D4}}\right)^{3/4}, \quad 
	R_{D4}^3 \equiv \pi g_s N l_s^3, \quad 
	f(u) \equiv 1-\left(\frac{u_\Lambda}{u}\right)^3,
\end{split}
\end{equation}
The string coupling $g_s$ is related to the $4+1$ Yang-Mills coupling by $4\pi^2 g_s l_s = g_5^2$. Here $x_4 \sim x_4 + 2\pi R$ with $R = \frac{2}{3} \left(\frac{R_{D4}^3}{u_\Lambda}\right)^{1/2}$ (by demanding smooth solution where the $x_4$ circle shrinks at $u=u_\Lambda$).

The thermodynamics of this system was studied in \cite{Aharony:2006da}.
The thermal solution is produced by analytically continuing the time direction to a circle of size $\beta$, $x_0 = i t$ $t \sim t+\beta$. On the gravity side, the analytically continued solution is the dominating solution for low temperatures. For high temperatures, one can take the same solutions, only with the $R$ and $\beta$ circles switched, so that now the temporal circle shrinks in the bulk. This last solution is a Euclidean black brane solution. The phase transition between the two happens at $\beta=2\pi R$ \cite{Aharony:2006da}.

Here we are interested in winding condensate solutions around the (analytically continued) thermal saddle \eqref{eq:witten_d4}. The solutions extend homogeneously on the $x^i$ directions but localize around $u=u_\Lambda$.
Defining $r^2 = \frac{4}{3} \left(\frac{R_{D4}^3}{u_\Lambda}\right)^{1/2} (u-u_\Lambda)$, $x_0 = i t$ and $\phi = \frac{x_4}{R}$ we have to leading order in $r^2$
\begin{equation}
\begin{split}
	ds^2 & = \left(\frac{u_\Lambda}{R_{D4}}\right)^{3/2} \left[ dt^2 + \sum_{i=1}^3 (dx^i)^2\right] + r^2 d\phi^2 + dr^2 + \left(\frac{R_{D4}^3}{u_\Lambda}\right)^{1/2} d\Omega_4^2.
\end{split}
\end{equation}
And at the next order in $r^2$
\begin{equation}
	G_{tt} = \left(\frac{u_\Lambda}{R_{D4}}\right)^{3/2}\left(1+\frac{3}{2} \frac{u-u_\Lambda}{u_\Lambda}\right) = \left(\frac{u_\Lambda}{R_{D4}}\right)^{3/2}\left(1+\frac{27 R}{16 R_{D4}^3} r^2\right).
\end{equation}
Comparing with \eqref{eq:conf_ansatz} we find the curvature scale $l^2 = \frac{16}{27} \frac{R_{D4}^3}{R}$.
In this model, the (confining) string tension is given by $T_{st} = \frac{1}{2\pi \alpha'} \left(\frac{u_\Lambda}{R_{D4}}\right)^{3/2} = \frac{4}{27 \pi \alpha'} (R_{D4}/R)^3$ and the mass gap (called the glueball mass) is of order $M_{gb} = 1/R$. Plugging inside \eqref{eq:conf_hag} with $d=2$ gives the leading correction to the holographic Hagedorn temperature
\begin{equation}\label{eq:witten_d4_hag}
\begin{split}
	\beta_c  = \sqrt{\frac{4\pi}{T_{st}}} - \frac{1}{\sqrt{2}}\frac{M_{gb}}{T_{st}}+ O(M_{gb}^2 T_{st}^{-3/2}),
\end{split}
\end{equation}
or in terms of the dimensionless ratio $\lambda_5/R$:
\begin{equation}
	\beta_c / R = \sqrt{108 \pi^3}\cdot \left(\lambda_5/R\right)^{-1/2}
            - \frac{27 \pi^2}{\sqrt{2} \left(\lambda_5/R\right)^{-1}}
            +O((\lambda_5/R)^{-3/2}).
\end{equation}
For a worldsheet derivation of \eqref{eq:witten_d4_hag}, see \cite{Bigazzi:2022gal}.

We found that this model admits a `winding-string gas' saddle for near-Hagedorn temperatures: a normalizable condensate of winding strings around Euclidean time $t$ that extends homogeneously on the $R^3$ directions (and $x^4$). The condensate is localized in the bulk around $u=u_\Lambda$ with a size of $L \sim \sqrt{l_s l} \sim l_s \cdot (T_{st} / M_{gb}^2)^{1/4}$.

The winding-string gas saddle is highly metastable, as it has free energy higher than both the thermal solution and (because we are above the phase transition) the black brane solution.
Note that the phase diagram is very similar to that of the \ads{3}, due to $2\pi R \iff \beta$ symmetry. Just like for the BTZ solution, the black brane solution has a tachyonic instability below the temperature
\begin{equation}
	\beta'_c = \frac{4\pi}{3} R_{D4} \left(\frac{R^2}{\alpha'}\right)^{1/3} = 2^{4/3} \pi \left(T_{st} M_{gb}^{-5}\right)^{1/3}.
\end{equation}
For temperatures slightly above it, there exists a condensate of winding $x^4$ strings homogeneous on $R^3\times S^1_\beta$, obtaining by interchanging the asymptotic circles for the winding-string gas saddle. It is possible to speculate a new kind of correspondence principle or a black brane/string transition. The black brane with $x^4$ winding condensate and the thermal solution with $t$-winding condensate are connected via a line of metastable saddles.

To understand the relation to pure Yang-Mills, it is useful to write the critical temperatures in terms of $R$ and the $4+1$ Yang-Mills coupling $\lambda_5$
\begin{equation}\label{eq:temps}
	\beta_c/R = \sqrt{108 \pi^3\frac{R}{\lambda_5}}, \quad \beta_{HP}/R = 2\pi, \quad \beta_c'/R = \frac{2}{3} \left(\pi^2 \frac{\lambda_5}{R}\right)^{1/3}.
\end{equation}
The semiclassical approximation is valid when $l \gg l_s \gg l_{pl}$ or $1 \ll \lambda_5 / R \ll N$. In this regime $\beta_c / R \ll 1 \ll \beta_c' / R$. In the opposite limit, $\lambda_5/R \ll 1$ the holographic theory at energies below $\sim 1/R$ is that of pure $3+1$ Yang-Mills. 
In \cite{Aharony:2006da} the phase diagram of the system was suggested for any $\lambda_5,R$. At $\lambda_5/R \ll 1$ and for low temperatures $\beta \gg R$ the theory is described by $3+1$ SYM. This theory has a first-order confinement/deconfinement close to its Hagedorn temperature $\beta_c \gg R$. As a result, we also expect a phase transition at high temperatures $\beta_c' \ll R$. The simplest scenario suggested in \cite{Aharony:2006da} was a single new dominating phase for temperatures $\beta_c' < \beta <\beta_c$. In such a scenario, this phase should have the unique property that it maps to itself when interchanging $\beta,R$.\footnote{In other words, the line of solutions is smooth at $\beta=2\pi R$ and the saddle there is mapped to itself under switching the circles. In more complicated scenarios, there are further phase transitions and phases that are mapped to each other under interchanging $\beta,R$.}

\begin{figure}[th]
	\centering
	\includegraphics[width=0.9 \textwidth]{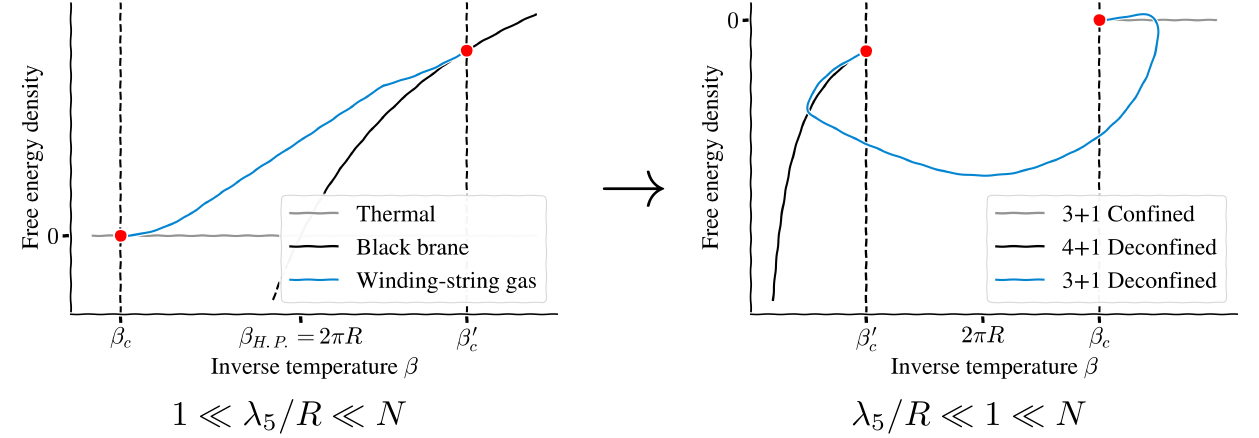}
	\caption{The phase diagram of the D4 Witten model as suggested by the analysis of \cite{Aharony:2006da}. 
	Left: The phase diagram when the bulk geometry is weakly curved $R \ll \lambda_5$. The two black lines represent the thermal and the black brane geometry, with a Hawking-Page transition at $\beta=2\pi R$. The two red dots represent the two critical temperatures  $\beta_c$, $\beta_c'$ above (below) which the thermal (black brane) saddle suffers from tachyonic instability. The blue line represents the `winding-string gas' saddle, which we suggest connects the two winding condensate solutions of each saddle.
	Right: In the regime $R \gg \lambda_5$ the low energy behavior is that of $3+1$ Yang-Mills, with a confinement/deconfinement transition at $\beta_c \gg R$. A similar transition happens for the `black brane' below some $\beta_c' \ll R$. We suggest that the dominating phase for $\beta_c'<\beta < \beta_c$ is nothing but the `winding-string gas' saddle.
	}
	\label{fig:witten_model}
\end{figure}

Going back to the gravity limit $1 \ll \lambda_5/R \ll N$. If indeed the winding-string gas saddle is the one that connects the two saddles between $\beta_c$ and $\beta_c'$, the simplest option would be that its continuation to the $\lambda_5/R \ll 1$ is the dominant phase for temperatures $\beta_c' < \beta <\beta_c$, see figure \ref{fig:witten_model}. Notice that because the winding-string gas phase breaks the winding of both circles, it is indeed invariant to switching $\beta,R$.
This picture is also supported by the analysis of \cite{Aharony:2005ew} which studied $2d$ SYM on a torus. In the weakly coupled regime, the authors found a dominating phase with non-zero holonomies on both circles.


\subsection{The D3 Witten model} \label{sec:d3_back}
It is possible to repeat the previous construction also for D3s compactified on a circle \cite{Witten:1998qj,Aharony:1999ti,Csaki:1998qr}. The holographic theory is now $3+1$ $\mathcal{N}=4$ $SU(N)$ Yang-Mills on $R^3 \times S^1$ with anti-periodic boundary conditions on the fermions. The gravity dual is the near-Horizon limit of $N$ D3 branes on $S^1$. After analytically continuing $x^0 = i t$ we find the thermal solution
\begin{equation}
\begin{split}
	ds^2 / l_{ads}^2 &= f(u) d\tau^2 + f^{-1}(u)du^2 + u^2 \left[ dt^2 + \sum_{i=1}^2 (dx^i)^2\right] + d\Omega^2_5,\\
	f(u) &= u^2 - \frac{u_0^4}{u^2}, \quad F_5 = 16 \pi  N \alpha'^2 \omega_5,
\end{split}
\end{equation}
with $l_{ads}^2 = \sqrt{4\pi g_s N} \alpha'$, $\tau \sim \tau + \pi/u_0$ (the $\tau$-circle closes at $u=u_0$) and the length of the asymptotic circle is $R=1/(2 u_0)$. There is a non-zero RR flux through the $S^5$, $\int_{S^5} \omega_5 = \pi^3$.
Close to $u=u_0$ the radial coordinate is $r = R ((u-u_0)/u_0)^{1/2}$ and 
\begin{equation}
	G_{tt} = R^2 u_0^2 \left(1 + 2 \frac{r^2}{l_{ads}^2}+...\right).
\end{equation}
Comparing with \eqref{eq:conf_ansatz} we find the curvature scale $l^2 = l_{ads}^2/2$ with $d=2$ (shrinking $S^1$). Following \eqref{eq:conf_hag} the leading correction to the holographic Hagedorn temperature is
\begin{equation}
	\beta_c = \sqrt{\frac{4\pi}{T_{st}}} \left(1 - (4\pi g_s N)^{-1/4} + O\left((4\pi g_s N)^{-1/2}\right)\right),
\end{equation}
with $T_{st} = R^2 u_0^2/(2\pi \alpha')$. The phase diagram of this model (when the gravity approximation is valid) is similar to the D4 case. A black brane solution exists by replacing the two Euclidean circles and dominates the canonical ensemble for high temperatures $\beta > 2\pi R$. The black brane is tachyonic for temperatures below (at leading $g_s N$ order)
\begin{equation}
	\beta_c' = \frac{\pi}{2^{3/2}} \frac{R}{l_s u_0} = \frac{\pi}{4} \sqrt{\frac{4\pi g_s N}{T_{st}}}.
\end{equation}
The same line of argument used for the D4 suggests a similar black brane/string transition. The black brane and the thermal saddle are connected between $\beta_c \le \beta \le \beta_c'$, with perturbative winding condensates being the near-critical description at each side.

\subsection{The Maldacena-Nu\~{n}ez background} \label{sec:mn_back}
In \cite{Maldacena:2000yy} Maldacena and Nu\~{n}ez (interpreting the solutions of \cite{Chamseddine:1997nm}) considered $N$ D5 branes on $R^4 \times S^2$, with a twisted normal bundle on $S^3$.
The holographic effective description is $5+1$ maximally supersymmetric YM on $R^4\times S^2$ with twisting of the $S^2$ using a gauge background for a $U(1)_R \subset SU(2)_R \subset SO(4)$ global symmetry. In terms of $R^4$ the twist preserves $\mathcal{N}=1$ supersymmetry. We denote the $5+1$ ’t~Hooft coupling $\lambda_6$ and the radius of the $S^2$ by $R$. 
The near horizon 10d metric and RR flux of the D5 system is given by
\begin{equation}
\begin{split}
	ds^2_{str} / \alpha' &= e^{\phi_D} \left( -(dx^0)^2 + \sum_{i=1}^3 (dx^i)^2 + N \left[ d\rho^2 + e^{2g(\rho)} d\Omega_2^2 + \frac{1}{4} \sum_{a=1}^3 (w_a - A_a)^2\right]\right)\\
	F_3 &= N \left[-\frac{1}{4} (w^1-A^1)\wedge(w^2-A^2)\wedge (w^3-A^3)+\frac{1}{4}\sum_{a=1}{3}F^a \wedge(w^a-A^a)\right],
\end{split}
\end{equation}
with
\begin{equation}
\begin{split}
		e^{2\phi_{D}} &= g_s^2 \frac{\sinh 2\rho}{2 e^{g}}=g_s^2\left(1+\frac{8}{9} \rho^2+O(\rho^4)\right),\\
		e^{2g} &= \rho \coth(2\rho) - \frac{\rho^2}{\sinh^2(2\rho)} - \frac{1}{4}.
\end{split}
\end{equation}
here $w_a$ label 1-forms on $S^3$, and $A_a$ is a smooth solution that asymptotes at large $\rho$ to the prescribed $S^2$ twist, and $F=dA$. In this geometry, the asymptotic $S^2$ shrinks at $\rho=0$.
This solution is weakly coupled and curved for $1 \ll g_s N \ll N$, or $1 \ll \lambda_6 /R^2 \ll N$. For $\lambda_6/R^2 \ll 1$ (in which gravity is highly curved) on the other hand the four-dimensional $\mathcal{N}=1$ theory decouples at energies below $1/R$ \cite{Maldacena:2000yy}.

The thermal ensemble is produced by analytically continuing $x_0 = i t/ l_s$ ($t \sim t+\beta$).
Close to $\rho=0$ the radial coordinate is $r = (\alpha' g_s N)^{1/2} \rho$, and the temporal (string) metric component is
\begin{equation}
	G_{tt} = e^{\phi_{D}} = g_s \left(1+ \frac{4}{9} \frac{r^2}{\alpha' g_s N} + O\left(\frac{r^4}{(\alpha'g_s N)^2}\right)\right).
\end{equation}
At $r=0$ the $S^3$ has no twist ($A_i$ is flat), and as long as the solution's size is smaller than $L \ll (\alpha'g_s N)^{1/2}$ we can approximate the geometry by $S_1 \times R^3 \times R^{3} \times S^3$ ($r$ being the radial direction in $R^3$).
Comparing with \eqref{eq:conf_ansatz}, we find `winding-string gas' saddles $\chi(r),\varphi(r)$ homogeneous on $R^3 \times S^3$ for near-Hagedorn temperatures.
In terms of section \ref{sec:gen_disc} $l = \frac{3}{2} (\alpha' g_s N)^{1/2}$ and $d=3$ (as we have a shrinking $S^2$), and the first correction to the Hagedorn temperature is \eqref{eq:conf_hag}
\begin{equation}
	\beta_c = \sqrt{\frac{4\pi}{T_{st}}} \left(1-(2 g_s N)^{-1/2} + O(N^{-1}) \right).
\end{equation}
with $T_{st} = \frac{1}{2\pi \alpha'} g_s$.\footnote{The high energy description of this model is in terms of little string theory, which has its own, different, Hagedorn temperature \cite{Kutasov:2000jp}. The latter appear at much higher temperatures $\beta_{H,LST} = 2\pi (\alpha' N)^{1/2} \ll \beta_c$ (in the case we are considering $g_s N \ll 1$). \label{fn:LST}}

Close to the Hagedorn temperature $\beta-\beta_c \ll \beta_c$ there exists a reliable winding-string gas saddle homogeneous on the $R^3$ (and the rest of the compact directions) and localize at the IR $\rho=0$ region with size $L \sim l_s (g_s N)^{1/4}$. The black brane-like solution, in which the thermal circle closes, and its thermodynamics were studied in \cite{Buchel:2001qi,Gubser:2001eg,Buchel:2001dg}.\footnote{For the thermodynamics of other supergravity duals of 4d confining gauge theories see \cite{Freedman:2000xb,Buchel:2003ah,Buchel:2007vy}
}
A black brane/string transition for this model suggests the black brane (with a small horizon radius) is connected to the winding-string gas saddle via a line of worldsheet CFTs.

\subsection{The Klebanov-Strassler conifold background} \label{sec:ks_back}
The Klebanov-Strassler geometry is given by \cite{Klebanov:2000hb}
\begin{equation} \label{eq:KS}
	ds^2 = h^{-1/2} \left(-(dx^0)^2 + \sum_{i=1}^3 (dx^i)^2\right) 
	+ h^{1/2}\left(
	\frac{\varepsilon^{4/3} K(\tau)}{6 K^3(\tau)} d\tau^2 + (...)
	\right)
\end{equation}
In $(...)$ we omitted the remaining 5 dimensions of the conifold geometry. For small $\tau$
\begin{equation}
\begin{split}
	K(\tau) &= (2/3)^{1/3} - \tau^2/(5 \cdot 2^{2/3}\cdot 3^{1/3}) + O(\tau^4)\\
	h(\tau) &= (g_s M \alpha')^2 2^{2/3} \varepsilon^{-8/3}(a_0 - 2^{-1/3} 3^{-4/3} \tau^2 + O(\tau^4)),
\end{split}
\end{equation}
and $a_0 \approx 0.71805$ \cite{PandoZayas:2003yb}. Here $M$ is the number of fractional $D3$ branes and $\varepsilon$ is the deformation parameter.
The background also includes RR and NS-NS fluxes, which are immaterial in the discussion of winding condensate.

The thermodynamics of Klebanov-Strassler was studied in \cite{Gubser:2001ri,Buchel:2001gw,Aharony:2007vg,Buchel:2018bzp}. Here we are interested only in the thermal solution, given by analytically continuing \eqref{eq:KS} by $x^0 = i l_s t$ ($t\sim t+\beta$). At $\tau=0$ an $S^2$ of the conifold geometry closes, and the geometry is approximately $R^4 \times R^3 \times S^3$.
Defining the radial coordinate $r=G_{\tau\tau}^{1/2} \tau$ we find \footnote{here the $x_i$ have units of length squared. When we write $G_{tt}$ we already multiplying by $\alpha'$ to make contact with \eqref{eq:conf_ansatz}.}
\begin{equation}
	G_{tt} = \frac{\varepsilon^{4/3}}{2^{1/3} a_0^{1/2} \ g_s M } \left(1 + \frac{1}{3 a_0^{3/2}} \frac{r^2}{g_s M \alpha'}\right).
\end{equation}
In terms of \eqref{eq:conf_ansatz} $d=3$ and a curvature scale $l^2 = 3 a_0^{3/2} g_s M \alpha'$. As a result, the leading correction to the Hagedorn temperature in this background is
\begin{equation}
	\beta_c = \sqrt{\frac{4\pi}{T_{st}}} \left(1 - \frac{3^{1/2}}{2^{3/2} a_0^{3/4}} \left(g_s M\right)^{-1/2} + O\left((g_s M)^{-1}\right) \right),
\end{equation}they
with the confining string tension $T_{st} = \frac{\varepsilon^{4/3}}{2^{4/3} a_0^{1/2} \pi} (\alpha' g_s M)^{-1}$.

\section{AdS-sized solutions for \titlemath{\ads{3}} with small RR flux} \label{sec:small_RR}
\subsection{The effective action} \label{sec:small_RR_anal}
In section \ref{sec:finite_RR} we concluded that the analysis for sub-AdS solutions breaks for small RR fluxes $1-\lambda^2 \sim \alpha'/l_{ads}^2$, where the solution's size reaches the AdS scale $L \sim l_{ads}$. For those cases, we can't neglect the backreaction of the winding mode on the supergravity fields \eqref{eq:deform}.
In this section, we turn to study the behavior of \ads{3} string stars for small RR fluxes, as well as the pure NS-NS case. To that end we define 
\begin{equation}
	1-\lambda^2 \equiv a \cdot \frac{\alpha'}{l_{ads}^2},
\end{equation}
and assume $0 \le a \ll l_{ads}^2/\alpha'$.\footnote{Notice that the quantization of the RR charge requires $\sqrt{1-\lambda^2} l_{ads}^2/\alpha'=\sqrt{a} l_{ads}/l_s$ to be integers. In our $l_{ads} \gg l_s$ limit it gives no restriction on the range of $a$.} Section \ref{sec:finite_RR} considered only near-Hagedorn temperatures $(R_0^2-R_H^2)/R_H^2 \sim l_s/l_{ads}$. 
For small RR flux, the important regime is the tighter $(R_0^2-R_H^2)/R_H^2 \sim \alpha'/l_{ads}^2$ (which we can call the very-near-Hagedorn regime).
We therefore define
\begin{equation}
	R_0^2 = R_H^2 + \hat m^2 \cdot \frac{\alpha'^2}{l_{ads}^2}.
\end{equation}
Solutions are expected to be valid as long as $R_0^2-R_H^2 \ll R_H^2$, or $\hat m^2 \ll l_{ads}^2/\alpha'$, which we will therefore assume. 

In the rest of the section, we will set the units to $l_{ads}=1$ for brevity. In terms of the new variables and at leading order in $l_s/l_{ads}$ the effective mass \eqref{eq:ads3_m2_eff} is
\begin{equation}
	m^2_\text{eff} = \hat m^2 + 2 a \ \sinh^2\rho.
\end{equation}
In the limit $\lambda \approx 1$ the RR flux terms in the action \eqref{eq:SUGRA_action_ads3} are subleading and can be neglected $\eta=0$. 
For the rest of the fields, it is useful to redefine by (in $l_{ads}=1$ units)
\begin{equation}
	\chi = \alpha' \hat \chi, \quad 
	\phi = \alpha' \hat \phi, \quad
	\varphi = \alpha' \hat \varphi, \quad
	\psi = \alpha' \hat \psi, \quad
	\zeta = \alpha' \hat \zeta.
\end{equation}
At leading order, we get the action 
\begin{align}
	I_\text{String star} -I_{\text{TAdS}_3} &= \frac{\beta \alpha'^2}{16\pi G_N} \cdot F,\label{eq:Delta_I_alpha}
\end{align}
with, upon integrating by parts, the normalized free energy (compared to that of thermal \ads{3})
\begin{equation}\label{eq:F_def}
\begin{split}
	F  = \int d\rho \ \text{vol} \  &\Bigg[
	(\hat\psi')^2 +(\hat\varphi')^2-4 (\hat\phi')^2 \\
	&-\frac{1}{2}\tanh^2\rho \ (\hat\zeta')^2-4(\hat\psi+\hat\varphi-\hat\zeta)^2
	+4\tanh\rho \ \hat\zeta'(\hat\psi+\hat\varphi-\hat\zeta)\\
	&+4\hat\phi'\left(
	\tanh\rho \  \hat\varphi + \coth\rho \  \hat\psi - \tanh\rho \  \hat\zeta\right)\\
	&+|\hat\chi'|^2+m^2_\text{eff}\rho \  |\hat\chi|^2 + 4 \left( \cosh^2\rho \ \hat\varphi -\sinh^2\rho\  (\hat\zeta-\hat\psi)\right) |\hat\chi|^2
	\Bigg],
\end{split}
\end{equation}
The equation of motion for the dilaton gives the first-order constraint (upon using the boundary conditions for $\hat \psi$)
\begin{equation} \label{eq:phi_const}
	2\hat\phi' = \tanh\rho \  \hat\varphi + \coth\rho \  \hat\psi - \tanh\rho \  \hat\zeta.
\end{equation}
Substituting \eqref{eq:phi_const} gives the equations of motion for the rest of the fields 
\begin{equation}\label{eq:large_k_eqs_alpha}
\begin{split}
	\hat \chi''+v(\rho) \hat \chi'&=(\hat m^2 + 2 a \ \sinh^2\rho) \hat \chi+4 \left(\cosh^2\rho \  \hat \varphi+\sinh^2\rho \ (\hat\psi-\hat\zeta)\right)\hat\chi,\\
	\hat\varphi''+v(\rho) \hat\varphi'&=
	2\tanh\rho \  \hat\zeta'
	+\tanh^2\rho \  (\hat\varphi-\hat\zeta)+\hat\psi
	-4(\hat\varphi+\hat\psi-\hat\zeta)
	+2 \cosh^2\rho \ |\hat\chi|^2,\\
	\hat\psi''+v(\rho) \hat\psi' &=
	2\tanh\rho \  \hat\zeta'
	+\hat\varphi-\hat\zeta+\coth^2\rho \  \hat\psi
	-4(\hat\varphi+\hat\psi-\hat\zeta)
	+2 \sinh^2\rho \ |\hat\chi|^2,\\
	\hat\zeta''+u(\rho)\hat\zeta'&=
	4\coth\rho \ (\hat\varphi'+\hat\psi')
	+2(\hat\varphi-\hat\zeta+\coth^2\rho \ \hat\psi) 
	+4 \cosh^2\rho \ |\hat\chi|^2,
\end{split}
\end{equation}
with
\begin{equation}
	v(\rho) = \tanh\rho +\coth\rho , \quad u(\rho) = 3\coth\rho -\tanh\rho .
\end{equation}
$a$ adds a potential to the linear equation for $\hat \chi$ and therefore for $a > 0$ we expect a normalizable solution with size $L \sim 1$. Assuming an $O(1)$ result (for $\hat m^2 \sim O(1)$) gives $|\chi|,|\varphi|,|\zeta| \sim \alpha' \ll 1$ which is self-consistent with our assumptions. The reader may worry that as the effective mass is now $m^2_\text{eff}(\rho) \sim O(1)$ (in $l_{ads}$ units), higher curvature corrections are now leading. In section \ref{sec:higher_curv} below we argue that the resulting mass shift is proportional to $\Delta m^2 \propto (1-\lambda^2)$. For finite $a$ this is a $O(\alpha')$ correction to the mass and therefore subleading.

In the extreme case $a=0$ ($\lambda=1$) we find the pure NS-NS background. In this case, the $\rho$ dependence in $m^2_\text{eff}$ is exactly canceled between the metric and the NS-NS flux, which results in a constant mass
\begin{equation}\label{eq:eff_m2_type_ii}
	m^2_\text{eff}(\rho) = \hat m^2 = \frac{1}{\alpha'^2}(R^2_0-R_H^2),
\end{equation}
equals to the flat space result \eqref{eq:m2_def}.
This property means that very close to the Hagedorn temperature strings winding around the thermal circle are light arbitrarily close to the boundary $\rho \gg 1$, although the proper length of the thermal circle diverges exponentially. Physically, this is due to the cancellation between the gravitational attraction (of AdS space) and the electric repulsion (or the non-zero $H_3$). This is related to the well-studied pure NS-NS \ads{3} long strings \cite{Maldacena:1998uz,Seiberg:1999xz}. These strings wind around the angular circle of Lorentzian \ads{3} with finite energy arbitrarily close to the boundary. Because asymptotically there is no difference between the two circles, the cancellation of the energy divergence at large $\rho$ is of the same origin.
As a result, for $a = 0$ the potential in \eqref{eq:large_k_eqs_alpha} is quadratic and not exponential. Nevertheless, we expect a normalizable solution for $\chi,\varphi,\zeta$ also in this case for high enough $\hat m^2$, with a generic $L \sim 1$. We present numerical solutions to this equation in section \ref{sec:small_RR_num} below.

\subsection{The \titlemath{\ads{3}} Hagedorn temperature}
To find the leading correction to the \ads{3} Hagedorn temperature, we should look for the maximal value of $\hat m^2$ (as a function of $a \ge 0$) for which there's a solution to the linearized \eqref{eq:large_k_eqs_alpha} equation for $\hat\chi$\footnote{At the linearized order the other fields in \eqref{eq:large_k_eqs_alpha} have no non-trivial solutions.}
\begin{equation}
	\hat \chi ''+ v(\rho) \, \hat \chi ' - \left(\hat m^2 +2a \ \sinh^2\rho \right) \hat \chi = 0.
\end{equation}
Denoting this critical value by $\hat m^2_c$, the first correction to the \ads{3} Hagedorn temperature is
\begin{equation}
	R_c^2(\lambda) = \frac{2}{k} + \frac{\hat m^2_c}{k^2} + O(k^{-3}),
\end{equation}
with $k=l_{ads}^2/\alpha'$.
For $a=0$ the (smooth) linear solution can be found analytically to be
\begin{equation}\label{eq:pure_linear_mode}
	\hat \chi_\text{lin}(\rho) = P_{\frac{-1+\sqrt{1+\hat m^2}}{2}}(\cosh(2\rho)),
\end{equation}
with $P_n(x)$ being the Legendre function \cite{Berkooz:2007fe}. For $\rho \gg 1$ and $\hat m^2 > -1$ we have $\hat \chi_\text{lin} \sim \exp(-\Delta_\chi \rho)$ with $\Delta_\chi = \Re(1-\sqrt{1+\hat m^2})$, and as $\hat \chi_\text{lin} \sim \rho \exp(-\rho)$ for $\hat m^2=-1$. The solution is thus non-normalizable for any $\hat m^2$, but it turns delta-function normalizable at $\hat m^2 = -1$ (at the Breitenlohner-Freedman bound). Therefore for $a=0$ the critical value is $\hat m^2_c = -1$, and the first correction to the pure NS-NS \ads{3} Hagedorn temperature is
\begin{equation}
	R_c^2(\lambda=1) = \frac{2}{k} - \frac{1}{k^{2}} + O(k^{-3}).
\end{equation}
Interestingly, for the type II pure NS-NS this is also the exact result $R_c^2 = 2/k - 1/k^2$ \cite{Berkooz:2007fe,Mertens:2014nca}.\footnote{I thank David Kutasov for pointing out that the first correction here is $O(k^{-2})$, and not $O(k^{-3/2})$ as found above \eqref{eq:finite_RR_R_c} and in \cite{Urbach:2022xzw}. It is his comment that led me to study the \ads{3} case.}
The fact that the zero-mode \eqref{eq:pure_linear_mode} is only delta-function normalizable is similar to flat space \cite{Atick:1988si}, and probably corresponds to the fact the Lorentzian spectrum, in this case, is continuous.

\begin{figure}[th]
	\centering
	\includegraphics[width=0.7\linewidth]{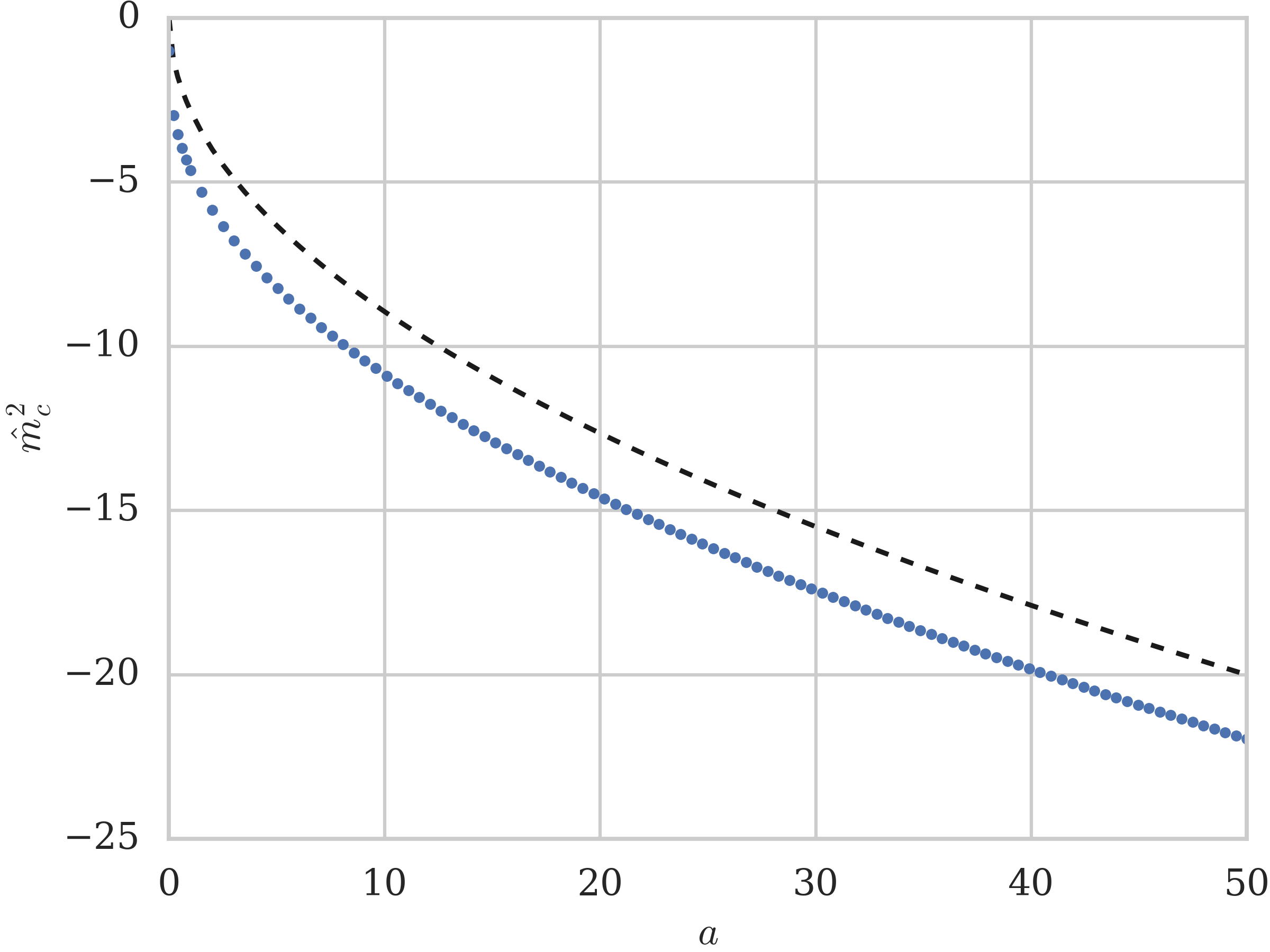}
	\caption{The value of the leading correction to the Hagedorn temperature $\hat m^2_c$ as a function of $a$ (blue). The $a \gg 1$ behavior from section \ref{sec:finite_RR} is plotted in a black dashed line.}
	\label{fig:m2s_alpha}
\end{figure}

For $a>0$ the linear equation has a potential and so we expect only discrete values of $\hat m^2$ for which exist normalizable solutions. We are interested in the `ground state' solution, with the largest $\hat m^2$. 
We weren't able to find an analytical solution in this case, but it is possible to find it numerically. In figure \ref{fig:m2s_alpha} we plotted $\hat m_c^2$ as a function of $a$. It can be seen from the figure that $\hat m^2$ decreases as $a$ grows. This is to be expected, as when $a \gg 1$ we expect from section \ref{sec:finite_RR} to find $\hat m^2_c = - 2^{3/2} a^{1/2}$ (by translating \eqref{eq:finite_RR_R_c} in terms of $\hat m^2$). This behavior is shown in the figure by a black dashed line.

\subsection{An argument against higher curvature corrections}\label{sec:higher_curv}
Higher curvature corrections always appear with non-negative powers of the string scale and therefore are naively subleading. For $d=2$, two operators still contribute at zeroth order in the string scale: $\mathcal{R}_{2} |\chi|^2$ and $g^{tt} F_{ij} F^{ij} |\chi|^2$. For the mixed background \eqref{eq:mixed_NS_NS_sol} their value is exactly a constant
\begin{equation}
	\mathcal{R}_{2} = -2,\quad g^{tt} F_{ij} F^{ij} = -4\lambda^2.
\end{equation}
We can write these terms in the Lagrangian as
\begin{equation}
	\mathcal{L}_\text{hige-curv} = g_1 \cdot \mathcal{R}_{2} |\chi|^2 + g_2 \cdot \frac{1}{2}g^{tt} F^{ij} |\chi|^2 = -2(g_1 + g_2 \lambda^2) |\chi|^2,
\end{equation}
for some unknown couplings $g_1,g_2 \sim O(1)$. 
Therefore the overall effect of these two terms is to shift the effective mass of $\chi$ by a constant $\Delta m^2 = -2(g_1 + g_2 \lambda^2)$.\footnote{Of course, both terms also induce a correction to the coupling of $\varphi$, $\zeta$ to $|\chi|^2$. Whenever the constant shift is subleading, so is the coupling.} As a result, the effective mass for the pure NS-NS case is ($k=l_{ads}^2/\alpha'$)
\begin{equation}\label{eq:Delta_m}
	m^2_\text{eff}(\rho) = k^2(R_0^2-R_H^2) + \Delta m^2 + O(1/k).
\end{equation}
We now turn specifically to the pure NS-NS theory $\lambda=1$, in which the worldsheet theory on thermal AdS is known \cite{Berkooz:2007fe,Rangamani:2007fz,Mertens:2014nca}.
The mass of the particle dictates the asymptotic behavior of its mode close to the boundary.
As such, it should match the \ads{3} $SL(2,R)$ Casimir of the corresponding worldsheet vertex operator. To find the exact worldsheet mass, we follow the discussion of \cite{Mertens:2014nca}, which uses the description of thermal \ads{3} as a specific coset of $SL(2,C)$ \cite{Hemming:2002kd}.
For a given $\beta$, a bosonic $SL(2,C)$ WZW primary with winding $\pm 1$ and $SL(2,C)$ Casimir $J^2 = s^2 + \frac{1}{4}$ (and $h_{int} =0$) has dimension
\begin{equation}
	h=\bar h = \frac{s^2 + \frac{1}{4}}{k-2} + \frac{k \beta^2}{16 \pi^2}
\end{equation}
We change variables to $R_0 = \frac{\beta}{2\pi}$, $-4m^2 = s^2 + \frac{1}{4}$ to get
\begin{equation}
	h=\bar h = -\frac{1}{4}\frac{m^2}{k-2} + \frac{k R_0^2}{4}.
\end{equation}
To write the type II expression we shift $k\mapsto k+2$ (only for the first term, as we keep the temperature in string units $k R_0^2$ constant) and find
\begin{equation}
	h=\bar h = -\frac{1}{4}\frac{m^2}{k} + \frac{k R_0^2}{4}=\frac{1}{2}.
\end{equation}
As a result, the mass of the winding vertex operator as a function of $R_0$ is
\begin{equation}
	m^2 = k^2 \left(R_0^2 - \frac{2}{k}\right).
\end{equation}
Comparing with \eqref{eq:Delta_m} we find that the higher curvature correction to the mass vanishes $\Delta m^2=0$, for $\lambda=1$. In other words $g_1=-g_2$, and (for any $\lambda$)
\begin{equation}
	\Delta m^2 = -2g_1(1- \lambda^2).
\end{equation}
Because in this section we consider only $1-\lambda^2 \sim 1/k$, this is a subleading effect.

\subsection{Numerical results} \label{sec:small_RR_num}
\begin{figure}[th]
	\centering
	\includegraphics[width=0.49\linewidth]{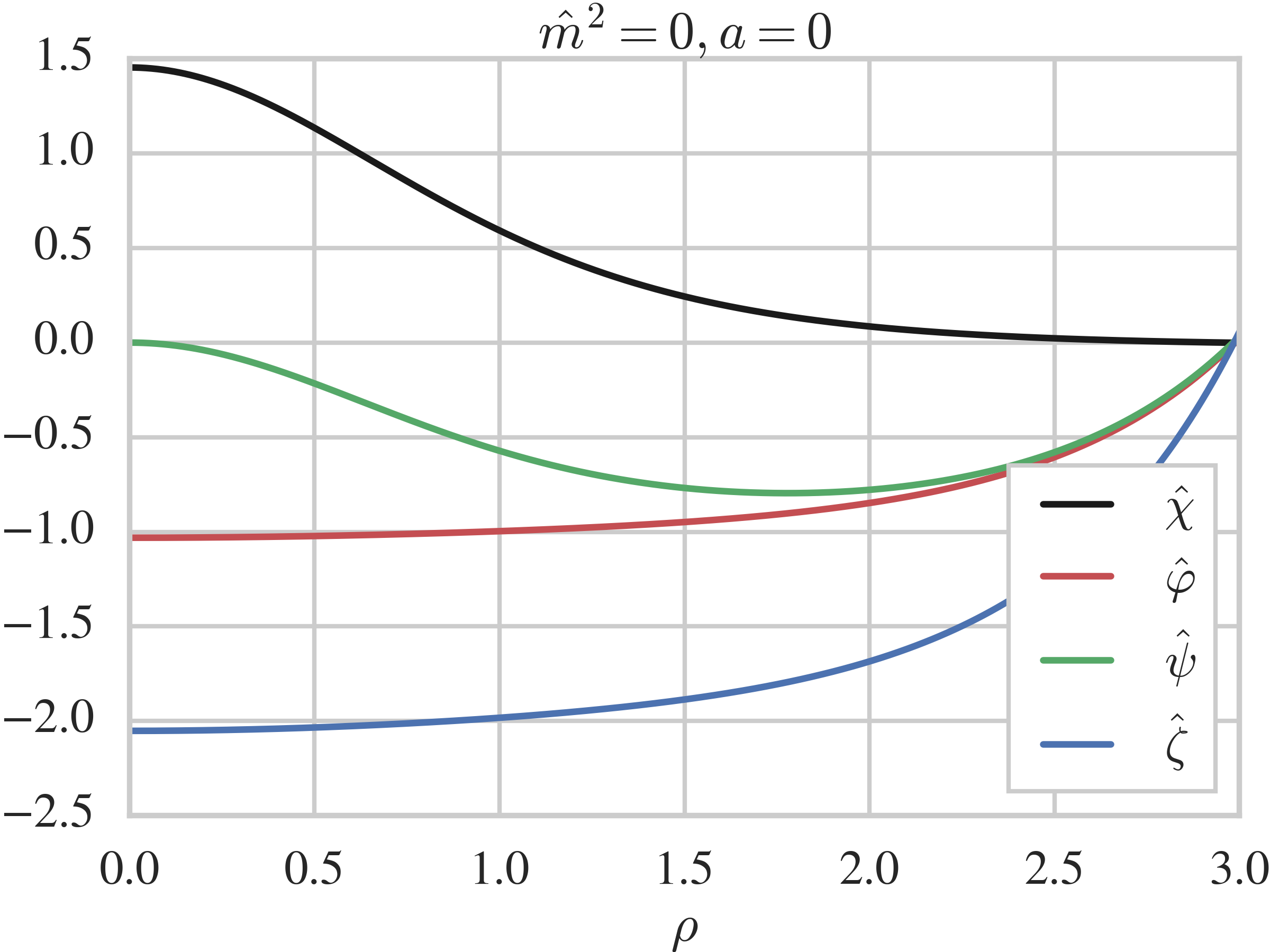}
	\includegraphics[width=0.49\linewidth]{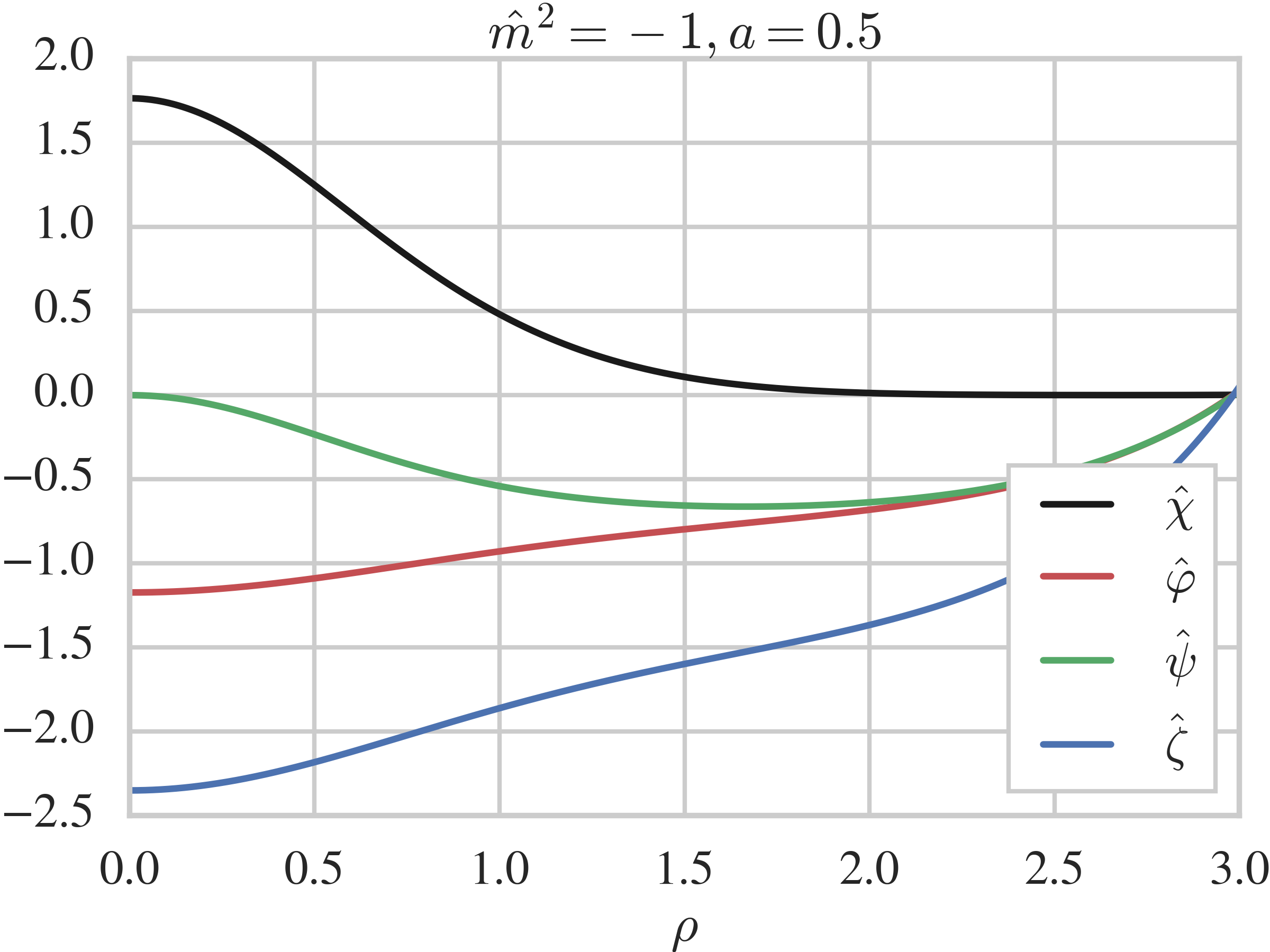}
	\caption{The two profiles of $\hat \chi$, $\hat \varphi$ and $\hat \zeta$ for temperature $\hat m^2 = 0$ in the pure NS-NS theory $a=0$ (left) and for $a=0.5$ (left).}
	\label{fig:example}
\end{figure}
	
\begin{figure}[th]
	\centering
	\includegraphics[width=0.49\linewidth]{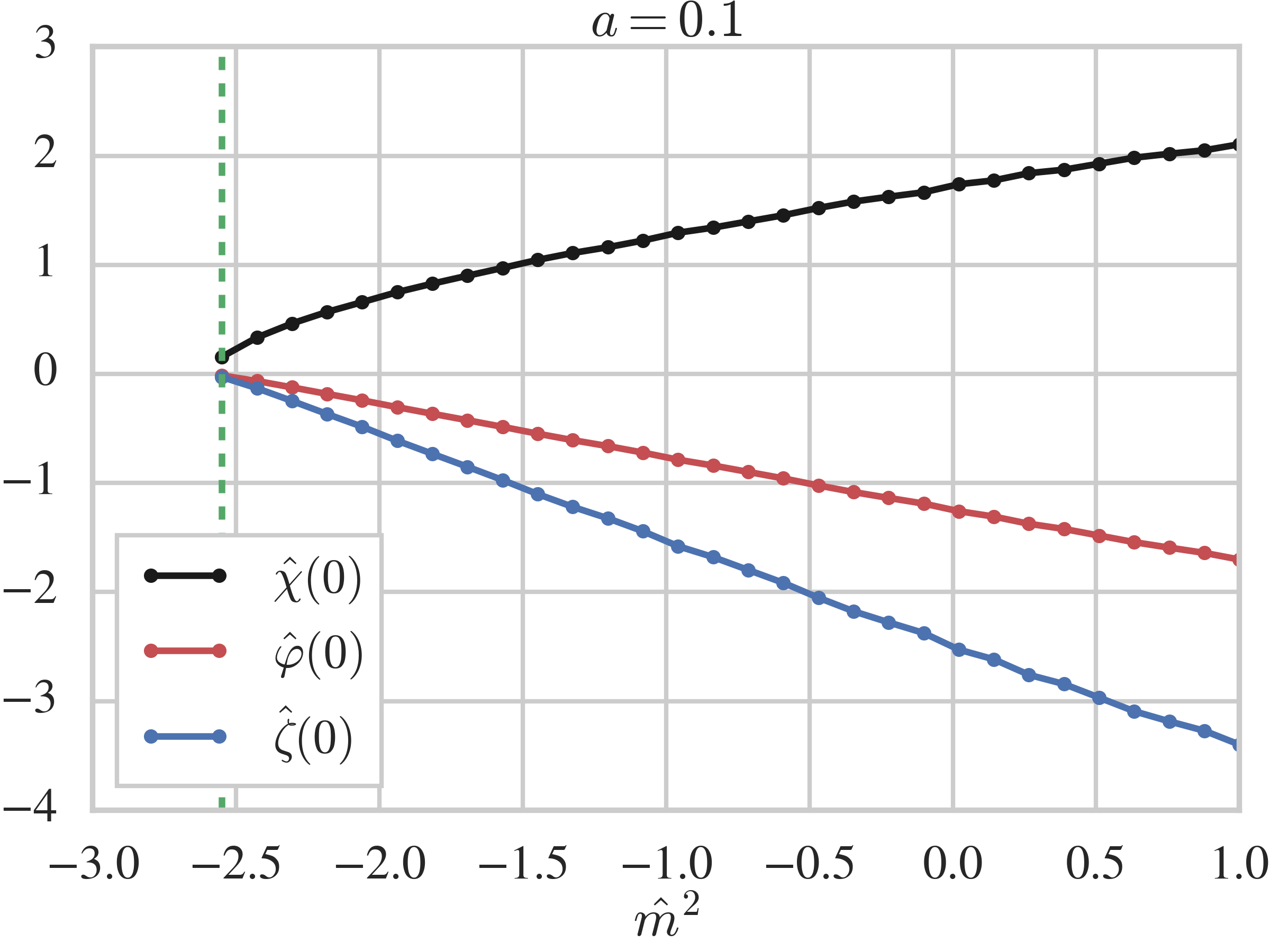}
	\includegraphics[width=0.49\linewidth]{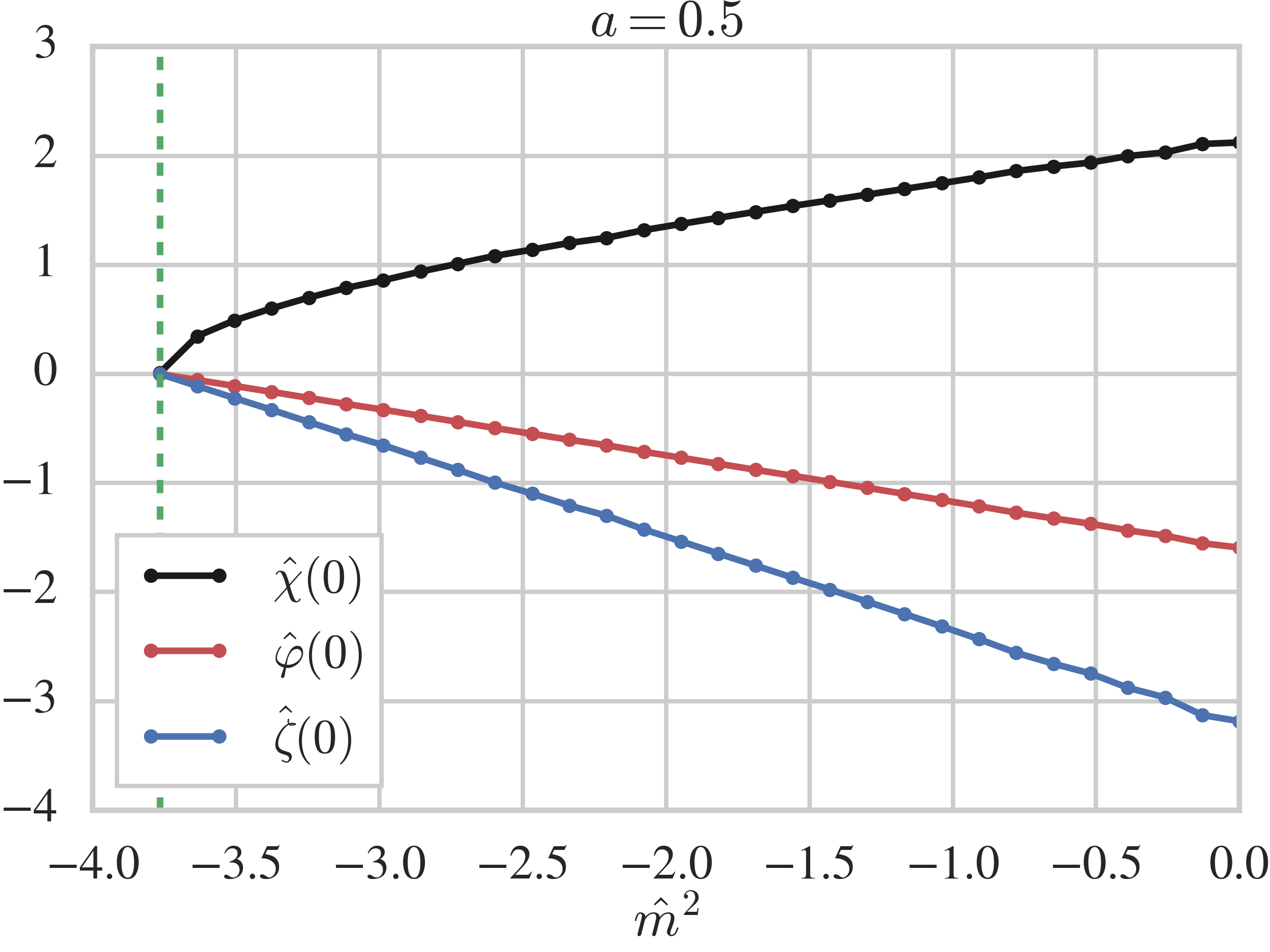}
	\caption{The value of $\hat\chi(0)$, $\hat\varphi(0)$ and $\hat\zeta(0)$ as a function of $\hat m^2$, for $a=0.1$ (left) and $a=0.5$ (right). The horizontal dashed green line indicates the AdS Hagedorn temperature $\hat m_c^2$.}
	\label{fig:chi0}
\end{figure}

\begin{figure}[th]
	\centering
	\includegraphics[width=0.49\linewidth]{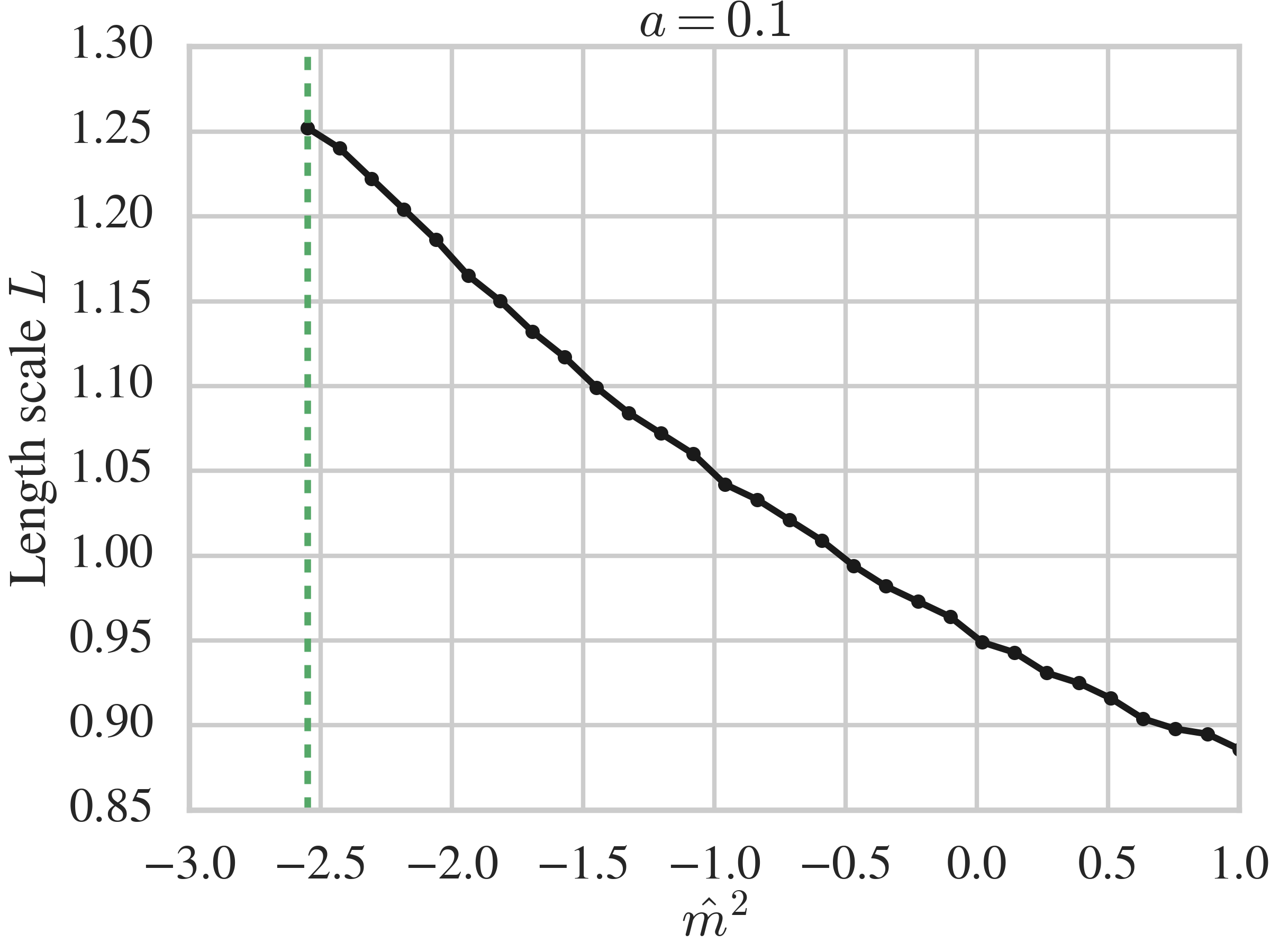}
	\includegraphics[width=0.49\linewidth]{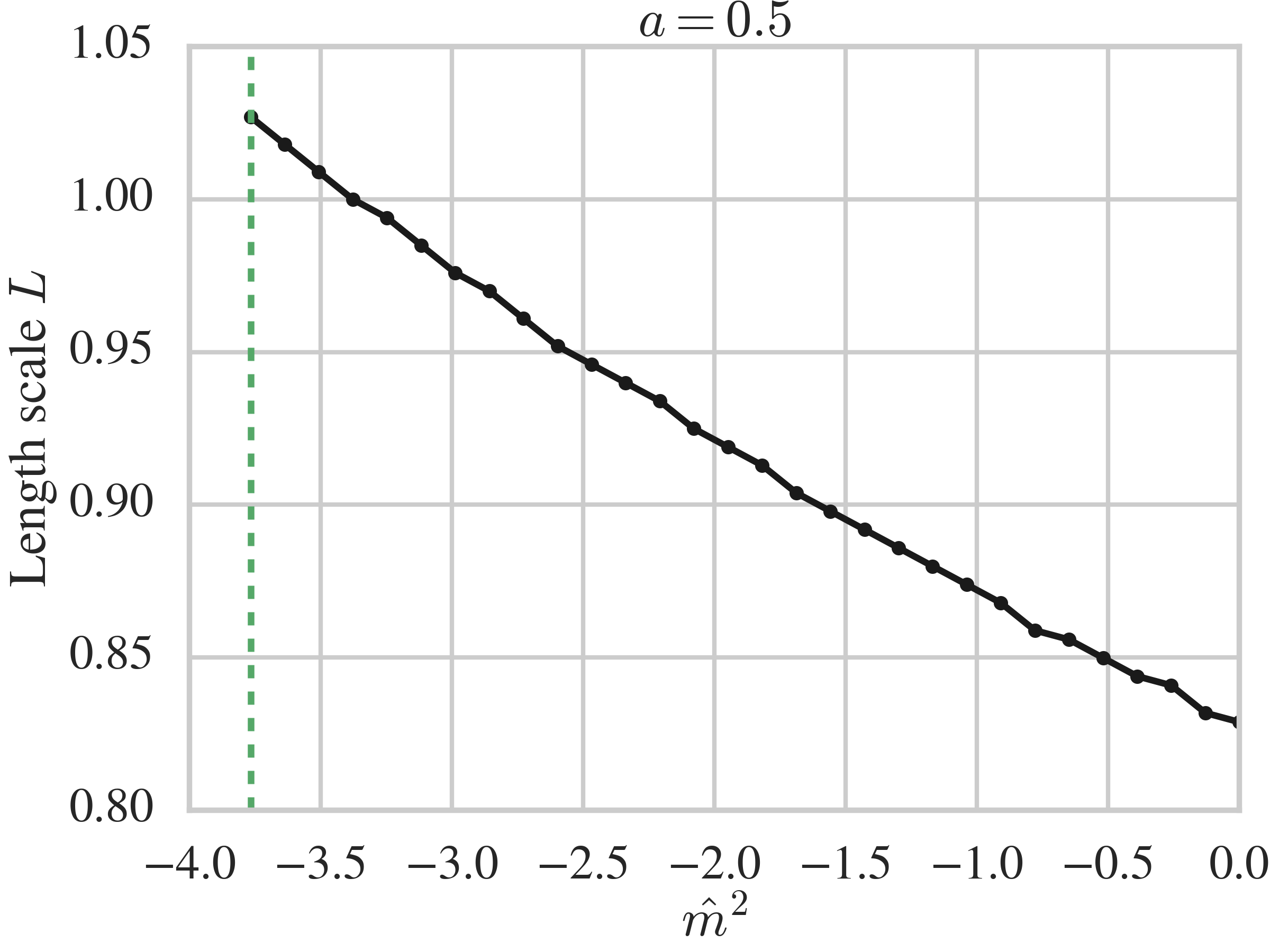}
	\caption{The size of $\hat \chi(\rho)$ as a function of $\hat m^2$, for $a=0.1$ (left) and $a=0.5$ (right). The size is defined as the radius $\rho$ in which the value of $\hat \chi(\rho)$ is decreased by a factor of $e$ compared to $\hat\chi(0)$. The horizontal dashed green line indicates the AdS Hagedorn temperature $\hat m_c^2$.}
	\label{fig:lengthscale}
\end{figure}

\begin{figure}[th]
	\centering
	\includegraphics[width=0.49\linewidth]{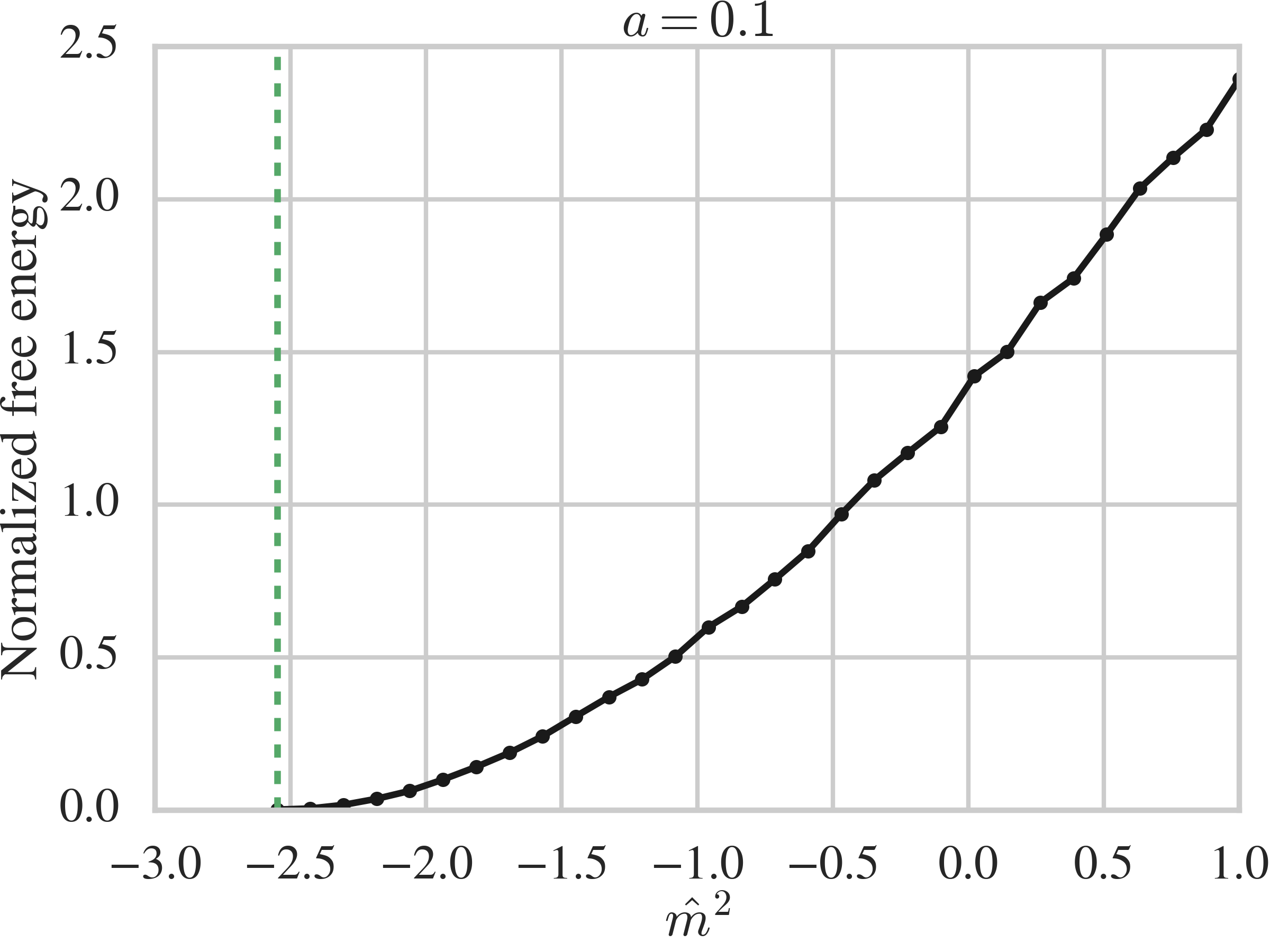}
	\includegraphics[width=0.49\linewidth]{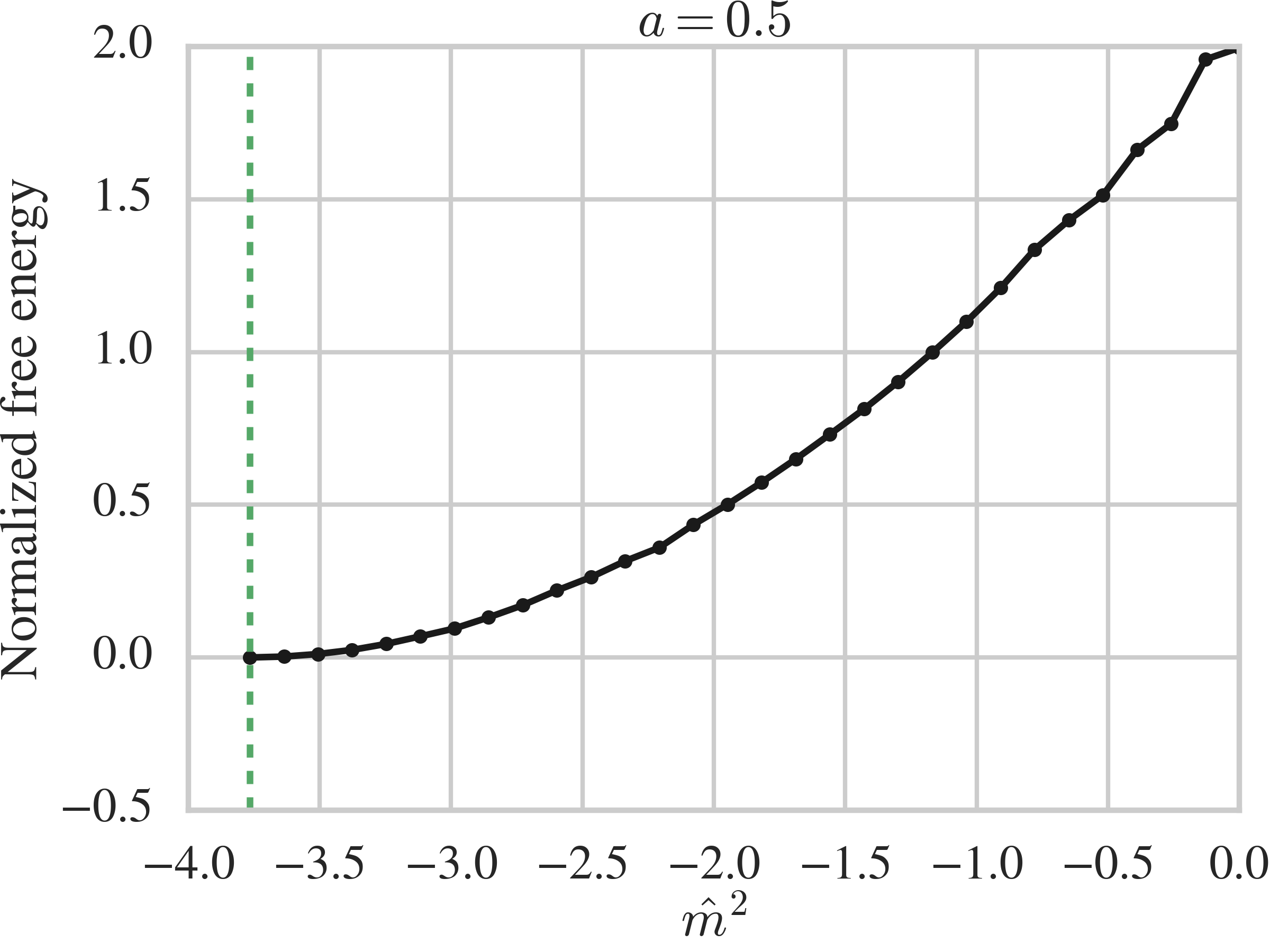}
	\caption{The string star normalized free energy as a function of $\hat m^2$, in the pure NS-NS theory $a=0.1$ (left) and $a=0.5$ (right). The horizontal dashed green line indicates the AdS Hagedorn temperature $\hat m_c^2$.}
	\label{fig:free_energy}
\end{figure}

To further study the profile of the AdS-sized string star solution we used numerical analysis to find the non-linear solutions for \eqref{eq:large_k_eqs_alpha} of $\hat \chi, \hat \varphi, \hat \psi$ and $\hat \zeta$ as a function of $a$ and $\hat m^2$. We set the boundary condition $\hat\psi(0)=0$ to avoid curvature singularity and for a smooth and normalizable solution 
\begin{equation}
	\hat \chi'(0) = \hat \varphi'(0) =\hat \psi'(0)= \hat\zeta'(0) = 0, \quad 
	\hat \chi(\infty) = \hat \varphi(\infty) =\hat\psi(\infty) = \hat\zeta(\infty) = 0.
\end{equation}
In practice, we chose $\rho_\text{min}=0.00001$ and $\rho_\text{max}=3$, and tried to satisfy $\psi(\rho_\text{min})=0$ with
\begin{equation}
	\hat \chi'(\rho_\text{min}) = \hat \varphi'(\rho_\text{min}) = \hat\psi'(\rho_\text{min}) = \hat\zeta'(\rho_\text{min}) = 0, \quad 
	\hat \chi(\rho_\text{max}) = \hat \varphi(\rho_\text{max}) = \hat\psi(\rho_\text{max}) = \hat\zeta(\rho_\text{max}) = 0,
\end{equation}
using the shooting method. That is, we optimized the initial values at $\rho_\text{min}$ to satisfy the boundary condition at $\rho_\text{max}$. Notice that the variable $\rho$ is already the dimensionless length in which the characteristic size of the solution should be $O(1)$.
Figure \ref{fig:example} shows the profile of the solution at $\hat m^2=0$ for the pure NS-NS system $a=0$ and for $a=0.5$.
We found that the near-Hagedorn solutions require higher and higher accuracy for smaller values of $a$. The reason is that $a$ controls the size of the potential for $\hat\chi$. In the extreme case of $a=0$, the linear equation for $\hat\chi$ has a continuous normalizable spectrum above the Hagedorn temperature. As a result, below we treat $a$ as a regulator for the simulation and plot $a=0.1$ instead of $a=0$.

As the solutions are monotonically decreasing functions of $x$, the initial value at $\rho=0$ sets the amplitude of the solution. Figure \ref{fig:chi0} shows the initial values $\hat\chi(0)$ (black), $\hat\varphi(0)$ (red) and $\hat\zeta(0)$ (blue) as a function of the temperature $\hat m^2$. 
For large $\hat m^2 \gg 1$ (lower temperatures) the solution amplitude seems to grow linearly with $\hat m^2$, or 
$|\chi| \sim (R_0^2-R_H^2)/R_H^2$. This is the same scaling found for (higher dimensional) flat space solutions \cite{Chen:2021dsw}. As in flat space, the solution is no longer reliable at $|\chi| \sim (R_0^2-R_H^2)/R_H^2 \sim 1$ or for $\hat m^2 \sim l_{ads}^2/\alpha'$. This is self-consistent, as we assumed $\hat m^2 \ll l_{ads}^2/\alpha'$ to derive \eqref{eq:large_k_eqs_alpha}. Notice that the temperature regime considered for finite RR flux $\hat m^2 \sim l_{ads}/l_s$ is inside the regime of validity.

As expected, the solution merges with the trivial solution $\hat \chi = \hat \varphi = \hat \psi = \hat \zeta = 0$ at the Hagedorn temperature $\hat m^2_c$ (drawn in green dashed line). Close to $\hat m^2_c$ the winding mode gets similar to the linear mode $\hat \chi_\text{lin}$ found numerically above, while $|\hat\varphi|,|\hat\psi|,|\hat\zeta|\ll |\hat\chi|$ are much smaller. This behavior is similar to the one found for higher dimensional AdS string stars \cite{Urbach:2022xzw} as well as for finite RR flux above (which is the large $a \gg 1$ limit of the simulation).

Figure \ref{fig:lengthscale} shows the size $L$ of the winding mode profile (in AdS units) as a function of the temperature. $L$ is defined as the value of $\rho$ at which the amplitude of $\hat\chi$ reaches $e^{-1} \hat\chi(0)$. The length grows with the temperature, as was found for finite $RR$ flux (including higher dimensions \cite{Urbach:2022xzw}). For finite $\hat m^2$ the size is at the AdS scale and bounded from above by the size of the linear solution at the Hagedorn temperature $\hat m^2=\hat m^2_c$. At lower temperatures, $\hat m^2 \gg 1$ the solution size reaches the string scale and the solutions are unreliable.

The normalized \ads{3} string star free energy $F$ \eqref{eq:F_def} is presented in figure \ref{fig:free_energy} as a function of the temperature. The free energy turns out to always be positive, which is not immediately clear from its form \eqref{eq:F_def}. The \ads{3} string star is therefore thermodynamically disfavored compared to the thermal \ads{3} solution. As we explain above, the solution is also non-perturbatively metastable due to the dominant BTZ solution. The free energy decreases with the temperature and goes to zero at the Hagedorn temperature when the solution becomes trivial.
General EFT arguments predict that close enough to the critical value $|\hat\chi| \sim (\hat m^2-\hat m^2_c)^{1/2}$ (and $|\hat\varphi|,|\hat\psi|,|\hat\zeta| \sim (\hat m^2-\hat m^2_c)^1$). As a result, we expect the Euclidean action to scale as
\begin{equation}
	I_\text{String star} -I_{\text{TAdS}_3} \sim \frac{\alpha'^2}{G_N} (\hat m^2 - \hat m^2_c)^2 \sim \frac{l_s}{G_N} \left(\frac{R_0-R_c}{R_c}\right)^2,
\end{equation}
where in the last line we moved back to general units for brevity.

\section*{Acknowledgments}

I would like to thank Micha Berkooz, Shai Chester, Rajesh Gopakumar, Indranil Halder, Daniel Jafferis, Yiyang Jia, Zohar Komargodski, Suman Kundu, Ohad Mamroud, Adar Sharon, Tal Sheaffer, Gabriel Wong and Xi Yin for useful discussions and comments. I especially thank Ofer Aharony and David Kutasov for their guidance and comments on the manuscript.\\
This work was partly funded by an Israel Science Foundation center for excellence grant (grant number 2289/18), by grant no. 2018068 from the United States-Israel Binational Science Foundation (BSF), by the Minerva foundation with funding from the Federal German Ministry for Education and Research, by the German Research Foundation through a German-Israeli Project Cooperation (DIP) grant ``Holography and the Swampland", and by a research grant from Martin Eisenstein.

\bibliographystyle{JHEP}
\bibliography{ref}

\providecommand{\href}[2]{#2}\begingroup\raggedright\begin{thebibliography}{10}

\bibitem{Urbach:2022xzw}
E.~Y. Urbach, \emph{{String stars in anti de Sitter space}},
  \href{https://doi.org/10.1007/JHEP04(2022)072}{\emph{JHEP} {\bfseries 04}
  (2022) 072} [\href{https://arxiv.org/abs/2202.06966}{{\ttfamily
  2202.06966}}].

\bibitem{Veneziano:1986zf}
G.~Veneziano, \emph{{A Stringy Nature Needs Just Two Constants}},
  \href{https://doi.org/10.1209/0295-5075/2/3/006}{\emph{EPL} {\bfseries 2}
  (1986) 199}.

\bibitem{Susskind:1993ws}
L.~Susskind, \emph{{Some speculations about black hole entropy in string
  theory}},  \href{https://arxiv.org/abs/hep-th/9309145}{{\ttfamily
  hep-th/9309145}}.

\bibitem{Horowitz:1996nw}
G.~T. Horowitz and J.~Polchinski, \emph{{A Correspondence principle for black
  holes and strings}},
  \href{https://doi.org/10.1103/PhysRevD.55.6189}{\emph{Phys. Rev. D}
  {\bfseries 55} (1997) 6189}
  [\href{https://arxiv.org/abs/hep-th/9612146}{{\ttfamily hep-th/9612146}}].

\bibitem{Horowitz:1997jc}
G.~T. Horowitz and J.~Polchinski, \emph{{Selfgravitating fundamental strings}},
  \href{https://doi.org/10.1103/PhysRevD.57.2557}{\emph{Phys. Rev. D}
  {\bfseries 57} (1998) 2557}
  [\href{https://arxiv.org/abs/hep-th/9707170}{{\ttfamily hep-th/9707170}}].

\bibitem{Damour:1999aw}
T.~Damour and G.~Veneziano, \emph{{Selfgravitating fundamental strings and
  black holes}},
  \href{https://doi.org/10.1016/S0550-3213(99)00596-9}{\emph{Nucl. Phys. B}
  {\bfseries 568} (2000) 93}
  [\href{https://arxiv.org/abs/hep-th/9907030}{{\ttfamily hep-th/9907030}}].

\bibitem{Khuri:1999ez}
R.~R. Khuri, \emph{{Selfgravitating strings and string / black hole
  correspondence}},
  \href{https://doi.org/10.1016/S0370-2693(99)01265-4}{\emph{Phys. Lett. B}
  {\bfseries 470} (1999) 73}
  [\href{https://arxiv.org/abs/hep-th/9910122}{{\ttfamily hep-th/9910122}}].

\bibitem{Kutasov:2005rr}
D.~Kutasov, \emph{{Accelerating branes and the string/black hole transition}},
  \href{https://arxiv.org/abs/hep-th/0509170}{{\ttfamily hep-th/0509170}}.

\bibitem{Giveon:2005jv}
A.~Giveon and D.~Kutasov, \emph{{The Charged black hole/string transition}},
  \href{https://doi.org/10.1088/1126-6708/2006/01/120}{\emph{JHEP} {\bfseries
  01} (2006) 120} [\href{https://arxiv.org/abs/hep-th/0510211}{{\ttfamily
  hep-th/0510211}}].

\bibitem{Chen:2021emg}
Y.~Chen and J.~Maldacena, \emph{{String scale black holes at large D}},
  \href{https://doi.org/10.1007/JHEP01(2022)095}{\emph{JHEP} {\bfseries 01}
  (2022) 095} [\href{https://arxiv.org/abs/2106.02169}{{\ttfamily
  2106.02169}}].

\bibitem{Brustein:2021cza}
R.~Brustein and Y.~Zigdon, \emph{{Black hole entropy sourced by string winding
  condensate}}, \href{https://doi.org/10.1007/JHEP10(2021)219}{\emph{JHEP}
  {\bfseries 10} (2021) 219}
  [\href{https://arxiv.org/abs/2107.09001}{{\ttfamily 2107.09001}}].

\bibitem{Chen:2021dsw}
Y.~Chen, J.~Maldacena and E.~Witten, \emph{{On the black hole/string
  transition}},  \href{https://arxiv.org/abs/2109.08563}{{\ttfamily
  2109.08563}}.

\bibitem{Balthazar:2022szl}
B.~Balthazar, J.~Chu and D.~Kutasov, \emph{{Winding Tachyons and Stringy Black
  Holes}},  \href{https://arxiv.org/abs/2204.00012}{{\ttfamily 2204.00012}}.

\bibitem{Bedroya:2022twb}
A.~Bedroya, \emph{{High energy scattering and string/black hole transition}},
  \href{https://arxiv.org/abs/2211.17162}{{\ttfamily 2211.17162}}.

\bibitem{Balthazar:2022hno}
B.~Balthazar, J.~Chu and D.~Kutasov, \emph{{On Small Black Holes in String
  Theory}},  \href{https://arxiv.org/abs/2210.12033}{{\ttfamily 2210.12033}}.

\bibitem{Atick:1988si}
J.~J. Atick and E.~Witten, \emph{{The Hagedorn Transition and the Number of
  Degrees of Freedom of String Theory}},
  \href{https://doi.org/10.1016/0550-3213(88)90151-4}{\emph{Nucl. Phys. B}
  {\bfseries 310} (1988) 291}.

\bibitem{Aharony:2003sx}
O.~Aharony, J.~Marsano, S.~Minwalla, K.~Papadodimas and M.~Van~Raamsdonk,
  \emph{{The Hagedorn - deconfinement phase transition in weakly coupled large
  N gauge theories}},
  \href{https://doi.org/10.4310/ATMP.2004.v8.n4.a1}{\emph{Adv. Theor. Math.
  Phys.} {\bfseries 8} (2004) 603}
  [\href{https://arxiv.org/abs/hep-th/0310285}{{\ttfamily hep-th/0310285}}].

\bibitem{Harmark:2021qma}
T.~Harmark and M.~Wilhelm, \emph{{Solving the Hagedorn temperature of
  AdS$_5$/CFT$_4$ via the Quantum Spectral Curve: Chemical potentials and
  deformations}},  \href{https://arxiv.org/abs/2109.09761}{{\ttfamily
  2109.09761}}.

\bibitem{Sundborg:1999ue}
B.~Sundborg, \emph{{The Hagedorn transition, deconfinement and N=4 SYM
  theory}}, \href{https://doi.org/10.1016/S0550-3213(00)00044-4}{\emph{Nucl.
  Phys. B} {\bfseries 573} (2000) 349}
  [\href{https://arxiv.org/abs/hep-th/9908001}{{\ttfamily hep-th/9908001}}].

\bibitem{Alvarez-Gaume:2005dvb}
L.~Alvarez-Gaume, C.~Gomez, H.~Liu and S.~Wadia, \emph{{Finite temperature
  effective action, AdS(5) black holes, and 1/N expansion}},
  \href{https://doi.org/10.1103/PhysRevD.71.124023}{\emph{Phys. Rev. D}
  {\bfseries 71} (2005) 124023}
  [\href{https://arxiv.org/abs/hep-th/0502227}{{\ttfamily hep-th/0502227}}].

\bibitem{Banados:1992wn}
M.~Banados, C.~Teitelboim and J.~Zanelli, \emph{{The Black hole in
  three-dimensional space-time}},
  \href{https://doi.org/10.1103/PhysRevLett.69.1849}{\emph{Phys. Rev. Lett.}
  {\bfseries 69} (1992) 1849}
  [\href{https://arxiv.org/abs/hep-th/9204099}{{\ttfamily hep-th/9204099}}].

\bibitem{Berkooz:2007fe}
M.~Berkooz, Z.~Komargodski and D.~Reichmann, \emph{{Thermal AdS(3), BTZ and
  competing winding modes condensation}},
  \href{https://doi.org/10.1088/1126-6708/2007/12/020}{\emph{JHEP} {\bfseries
  12} (2007) 020} [\href{https://arxiv.org/abs/0706.0610}{{\ttfamily
  0706.0610}}].

\bibitem{Lin:2007gi}
F.-L. Lin, T.~Matsuo and D.~Tomino, \emph{{Hagedorn Strings and Correspondence
  Principle in AdS(3)}},
  \href{https://doi.org/10.1088/1126-6708/2007/09/042}{\emph{JHEP} {\bfseries
  09} (2007) 042} [\href{https://arxiv.org/abs/0705.4514}{{\ttfamily
  0705.4514}}].

\bibitem{Rangamani:2007fz}
M.~Rangamani and S.~F. Ross, \emph{{Winding tachyons in BTZ}},
  \href{https://doi.org/10.1103/PhysRevD.77.026010}{\emph{Phys. Rev. D}
  {\bfseries 77} (2008) 026010}
  [\href{https://arxiv.org/abs/0706.0663}{{\ttfamily 0706.0663}}].

\bibitem{Jafferis:2021ywg}
D.~L. Jafferis and E.~Schneider, \emph{{Stringy ER = EPR}},
  \href{https://doi.org/10.1007/JHEP10(2022)195}{\emph{JHEP} {\bfseries 10}
  (2022) 195} [\href{https://arxiv.org/abs/2104.07233}{{\ttfamily
  2104.07233}}].

\bibitem{Halder:2022ykw}
I.~Halder, D.~L. Jafferis and D.~Kolchmeyer, \emph{{A duality in string theory
  on AdS$_3$}},  \href{https://arxiv.org/abs/2208.00016}{{\ttfamily
  2208.00016}}.

\bibitem{Maldacena:1997re}
J.~M. Maldacena, \emph{{The Large N limit of superconformal field theories and
  supergravity}}, \href{https://doi.org/10.1023/A:1026654312961}{\emph{Adv.
  Theor. Math. Phys.} {\bfseries 2} (1998) 231}
  [\href{https://arxiv.org/abs/hep-th/9711200}{{\ttfamily hep-th/9711200}}].

\bibitem{Giveon:1998ns}
A.~Giveon, D.~Kutasov and N.~Seiberg, \emph{{Comments on string theory on
  AdS(3)}}, \href{https://doi.org/10.4310/ATMP.1998.v2.n4.a3}{\emph{Adv. Theor.
  Math. Phys.} {\bfseries 2} (1998) 733}
  [\href{https://arxiv.org/abs/hep-th/9806194}{{\ttfamily hep-th/9806194}}].

\bibitem{Kutasov:1999xu}
D.~Kutasov and N.~Seiberg, \emph{{More comments on string theory on AdS(3)}},
  \href{https://doi.org/10.1088/1126-6708/1999/04/008}{\emph{JHEP} {\bfseries
  04} (1999) 008} [\href{https://arxiv.org/abs/hep-th/9903219}{{\ttfamily
  hep-th/9903219}}].

\bibitem{David:2002wn}
J.~R. David, G.~Mandal and S.~R. Wadia, \emph{{Microscopic formulation of black
  holes in string theory}},
  \href{https://doi.org/10.1016/S0370-1573(02)00271-5}{\emph{Phys. Rept.}
  {\bfseries 369} (2002) 549}
  [\href{https://arxiv.org/abs/hep-th/0203048}{{\ttfamily hep-th/0203048}}].

\bibitem{Cho:2018nfn}
M.~Cho, S.~Collier and X.~Yin, \emph{{Strings in Ramond-Ramond Backgrounds from
  the Neveu-Schwarz-Ramond Formalism}},
  \href{https://doi.org/10.1007/JHEP12(2020)123}{\emph{JHEP} {\bfseries 12}
  (2020) 123} [\href{https://arxiv.org/abs/1811.00032}{{\ttfamily
  1811.00032}}].

\bibitem{Maldacena:1998uz}
J.~M. Maldacena, J.~Michelson and A.~Strominger, \emph{{Anti-de Sitter
  fragmentation}},
  \href{https://doi.org/10.1088/1126-6708/1999/02/011}{\emph{JHEP} {\bfseries
  02} (1999) 011} [\href{https://arxiv.org/abs/hep-th/9812073}{{\ttfamily
  hep-th/9812073}}].

\bibitem{Seiberg:1999xz}
N.~Seiberg and E.~Witten, \emph{{The D1 / D5 system and singular CFT}},
  \href{https://doi.org/10.1088/1126-6708/1999/04/017}{\emph{JHEP} {\bfseries
  04} (1999) 017} [\href{https://arxiv.org/abs/hep-th/9903224}{{\ttfamily
  hep-th/9903224}}].

\bibitem{Witten:1998qj}
E.~Witten, \emph{{Anti-de Sitter space and holography}},
  \href{https://doi.org/10.4310/ATMP.1998.v2.n2.a2}{\emph{Adv. Theor. Math.
  Phys.} {\bfseries 2} (1998) 253}
  [\href{https://arxiv.org/abs/hep-th/9802150}{{\ttfamily hep-th/9802150}}].

\bibitem{Maldacena:2000yy}
J.~M. Maldacena and C.~Nunez, \emph{{Towards the large N limit of pure N=1
  superYang-Mills}},
  \href{https://doi.org/10.1103/PhysRevLett.86.588}{\emph{Phys. Rev. Lett.}
  {\bfseries 86} (2001) 588}
  [\href{https://arxiv.org/abs/hep-th/0008001}{{\ttfamily hep-th/0008001}}].

\bibitem{Klebanov:2000hb}
I.~R. Klebanov and M.~J. Strassler, \emph{{Supergravity and a confining gauge
  theory: Duality cascades and chi SB resolution of naked singularities}},
  \href{https://doi.org/10.1088/1126-6708/2000/08/052}{\emph{JHEP} {\bfseries
  08} (2000) 052} [\href{https://arxiv.org/abs/hep-th/0007191}{{\ttfamily
  hep-th/0007191}}].

\bibitem{Freedman:2000xb}
D.~Z. Freedman and J.~A. Minahan, \emph{{Finite temperature effects in the
  supergravity dual of the N=1* gauge theory}},
  \href{https://doi.org/10.1088/1126-6708/2001/01/036}{\emph{JHEP} {\bfseries
  01} (2001) 036} [\href{https://arxiv.org/abs/hep-th/0007250}{{\ttfamily
  hep-th/0007250}}].

\bibitem{Gubser:2001ri}
S.~S. Gubser, C.~P. Herzog, I.~R. Klebanov and A.~A. Tseytlin,
  \emph{{Restoration of chiral symmetry: A Supergravity perspective}},
  \href{https://doi.org/10.1088/1126-6708/2001/05/028}{\emph{JHEP} {\bfseries
  05} (2001) 028} [\href{https://arxiv.org/abs/hep-th/0102172}{{\ttfamily
  hep-th/0102172}}].

\bibitem{Buchel:2001gw}
A.~Buchel, C.~P. Herzog, I.~R. Klebanov, L.~A. Pando~Zayas and A.~A. Tseytlin,
  \emph{{Nonextremal gravity duals for fractional D-3 branes on the conifold}},
  \href{https://doi.org/10.1088/1126-6708/2001/04/033}{\emph{JHEP} {\bfseries
  04} (2001) 033} [\href{https://arxiv.org/abs/hep-th/0102105}{{\ttfamily
  hep-th/0102105}}].

\bibitem{Buchel:2001qi}
A.~Buchel and A.~R. Frey, \emph{{Comments on supergravity dual of pure N=1
  superYang-Mills theory with unbroken chiral symmetry}},
  \href{https://doi.org/10.1103/PhysRevD.64.064007}{\emph{Phys. Rev. D}
  {\bfseries 64} (2001) 064007}
  [\href{https://arxiv.org/abs/hep-th/0103022}{{\ttfamily hep-th/0103022}}].

\bibitem{Gubser:2001eg}
S.~S. Gubser, A.~A. Tseytlin and M.~S. Volkov, \emph{{NonAbelian 4-d black
  holes, wrapped five-branes, and their dual descriptions}},
  \href{https://doi.org/10.1088/1126-6708/2001/09/017}{\emph{JHEP} {\bfseries
  09} (2001) 017} [\href{https://arxiv.org/abs/hep-th/0108205}{{\ttfamily
  hep-th/0108205}}].

\bibitem{Buchel:2001dg}
A.~Buchel, \emph{{On the thermodynamic instability of LST}},
  \href{https://arxiv.org/abs/hep-th/0107102}{{\ttfamily hep-th/0107102}}.

\bibitem{Buchel:2003ah}
A.~Buchel and J.~T. Liu, \emph{{Thermodynamics of the N=2* flow}},
  \href{https://doi.org/10.1088/1126-6708/2003/11/031}{\emph{JHEP} {\bfseries
  11} (2003) 031} [\href{https://arxiv.org/abs/hep-th/0305064}{{\ttfamily
  hep-th/0305064}}].

\bibitem{Aharony:2006da}
O.~Aharony, J.~Sonnenschein and S.~Yankielowicz, \emph{{A Holographic model of
  deconfinement and chiral symmetry restoration}},
  \href{https://doi.org/10.1016/j.aop.2006.11.002}{\emph{Annals Phys.}
  {\bfseries 322} (2007) 1420}
  [\href{https://arxiv.org/abs/hep-th/0604161}{{\ttfamily hep-th/0604161}}].

\bibitem{Aharony:2007vg}
O.~Aharony, A.~Buchel and P.~Kerner, \emph{{The Black hole in the throat:
  Thermodynamics of strongly coupled cascading gauge theories}},
  \href{https://doi.org/10.1103/PhysRevD.76.086005}{\emph{Phys. Rev. D}
  {\bfseries 76} (2007) 086005}
  [\href{https://arxiv.org/abs/0706.1768}{{\ttfamily 0706.1768}}].

\bibitem{Buchel:2007vy}
A.~Buchel, S.~Deakin, P.~Kerner and J.~T. Liu, \emph{{Thermodynamics of the
  N=2* strongly coupled plasma}},
  \href{https://doi.org/10.1016/j.nuclphysb.2007.06.019}{\emph{Nucl. Phys. B}
  {\bfseries 784} (2007) 72}
  [\href{https://arxiv.org/abs/hep-th/0701142}{{\ttfamily hep-th/0701142}}].

\bibitem{Buchel:2018bzp}
A.~Buchel, \emph{{Klebanov-Strassler black hole}},
  \href{https://doi.org/10.1007/JHEP01(2019)207}{\emph{JHEP} {\bfseries 01}
  (2019) 207} [\href{https://arxiv.org/abs/1809.08484}{{\ttfamily
  1809.08484}}].

\bibitem{Maldacena:2000hw}
J.~M. Maldacena and H.~Ooguri, \emph{{Strings in AdS(3) and SL(2,R) WZW model
  1.: The Spectrum}}, \href{https://doi.org/10.1063/1.1377273}{\emph{J. Math.
  Phys.} {\bfseries 42} (2001) 2929}
  [\href{https://arxiv.org/abs/hep-th/0001053}{{\ttfamily hep-th/0001053}}].

\bibitem{halder_private}
N.~Agia, I.~Halder and D.~L. Jafferis. upcoming work.

\bibitem{Polchinski:1998rr}
J.~Polchinski, \emph{{String theory. Vol. 2: Superstring theory and beyond}},
  Cambridge Monographs on Mathematical Physics. Cambridge University Press, 12,
  2007,
  \href{https://doi.org/10.1017/CBO9780511618123}{10.1017/CBO9780511618123}.

\bibitem{Cavaglia:2021eqr}
A.~Cavagli\`a, N.~Gromov, B.~Stefa\'nski, Jr. and A.~Torrielli, \emph{{Quantum
  Spectral Curve for AdS$_{3}$/CFT$_{2}$: a proposal}},
  \href{https://doi.org/10.1007/JHEP12(2021)048}{\emph{JHEP} {\bfseries 12}
  (2021) 048} [\href{https://arxiv.org/abs/2109.05500}{{\ttfamily
  2109.05500}}].

\bibitem{Ekhammar:2021pys}
S.~Ekhammar and D.~Volin, \emph{{Monodromy bootstrap for SU(2|2) quantum
  spectral curves: from Hubbard model to AdS$_{3}$/CFT$_{2}$}},
  \href{https://doi.org/10.1007/JHEP03(2022)192}{\emph{JHEP} {\bfseries 03}
  (2022) 192} [\href{https://arxiv.org/abs/2109.06164}{{\ttfamily
  2109.06164}}].

\bibitem{Cavaglia:2022xld}
A.~Cavagli\`a, S.~Ekhammar, N.~Gromov and P.~Ryan, \emph{{Exploring the Quantum
  Spectral Curve for AdS${}_3$/CFT${}_2$}},
  \href{https://arxiv.org/abs/2211.07810}{{\ttfamily 2211.07810}}.

\bibitem{Frolov:2021bwp}
S.~Frolov and A.~Sfondrini, \emph{{Mirror thermodynamic Bethe ansatz for
  AdS3/CFT2}}, \href{https://doi.org/10.1007/JHEP03(2022)138}{\emph{JHEP}
  {\bfseries 03} (2022) 138}
  [\href{https://arxiv.org/abs/2112.08898}{{\ttfamily 2112.08898}}].

\bibitem{Witten:1998zw}
E.~Witten, \emph{{Anti-de Sitter space, thermal phase transition, and
  confinement in gauge theories}},
  \href{https://doi.org/10.4310/ATMP.1998.v2.n3.a3}{\emph{Adv. Theor. Math.
  Phys.} {\bfseries 2} (1998) 505}
  [\href{https://arxiv.org/abs/hep-th/9803131}{{\ttfamily hep-th/9803131}}].

\bibitem{Bigazzi:2022gal}
F.~Bigazzi, T.~Canneti and A.~L. Cotrone, \emph{{On the Hagedorn temperature in
  holographic confining gauge theories}},
  \href{https://doi.org/10.1007/JHEP01(2023)034}{\emph{JHEP} {\bfseries 01}
  (2023) 034} [\href{https://arxiv.org/abs/2210.09893}{{\ttfamily
  2210.09893}}].

\bibitem{Aharony:2005ew}
O.~Aharony, J.~Marsano, S.~Minwalla, K.~Papadodimas, M.~Van~Raamsdonk and
  T.~Wiseman, \emph{{The Phase structure of low dimensional large N gauge
  theories on Tori}},
  \href{https://doi.org/10.1088/1126-6708/2006/01/140}{\emph{JHEP} {\bfseries
  01} (2006) 140} [\href{https://arxiv.org/abs/hep-th/0508077}{{\ttfamily
  hep-th/0508077}}].

\bibitem{Aharony:1999ti}
O.~Aharony, S.~S. Gubser, J.~M. Maldacena, H.~Ooguri and Y.~Oz, \emph{{Large N
  field theories, string theory and gravity}},
  \href{https://doi.org/10.1016/S0370-1573(99)00083-6}{\emph{Phys. Rept.}
  {\bfseries 323} (2000) 183}
  [\href{https://arxiv.org/abs/hep-th/9905111}{{\ttfamily hep-th/9905111}}].

\bibitem{Csaki:1998qr}
C.~Csaki, H.~Ooguri, Y.~Oz and J.~Terning, \emph{{Glueball mass spectrum from
  supergravity}},
  \href{https://doi.org/10.1088/1126-6708/1999/01/017}{\emph{JHEP} {\bfseries
  01} (1999) 017} [\href{https://arxiv.org/abs/hep-th/9806021}{{\ttfamily
  hep-th/9806021}}].

\bibitem{Chamseddine:1997nm}
A.~H. Chamseddine and M.~S. Volkov, \emph{{NonAbelian BPS monopoles in N=4
  gauged supergravity}},
  \href{https://doi.org/10.1103/PhysRevLett.79.3343}{\emph{Phys. Rev. Lett.}
  {\bfseries 79} (1997) 3343}
  [\href{https://arxiv.org/abs/hep-th/9707176}{{\ttfamily hep-th/9707176}}].

\bibitem{Kutasov:2000jp}
D.~Kutasov and D.~A. Sahakyan, \emph{{Comments on the thermodynamics of little
  string theory}},
  \href{https://doi.org/10.1088/1126-6708/2001/02/021}{\emph{JHEP} {\bfseries
  02} (2001) 021} [\href{https://arxiv.org/abs/hep-th/0012258}{{\ttfamily
  hep-th/0012258}}].

\bibitem{PandoZayas:2003yb}
L.~A. Pando~Zayas, J.~Sonnenschein and D.~Vaman, \emph{{Regge trajectories
  revisited in the gauge / string correspondence}},
  \href{https://doi.org/10.1016/j.nuclphysb.2003.12.006}{\emph{Nucl. Phys. B}
  {\bfseries 682} (2004) 3}
  [\href{https://arxiv.org/abs/hep-th/0311190}{{\ttfamily hep-th/0311190}}].

\bibitem{Mertens:2014nca}
T.~G. Mertens, H.~Verschelde and V.~I. Zakharov, \emph{{The thermal scalar and
  random walks in $AdS_3$ and $BTZ$}},
  \href{https://doi.org/10.1007/JHEP06(2014)156}{\emph{JHEP} {\bfseries 06}
  (2014) 156} [\href{https://arxiv.org/abs/1402.2808}{{\ttfamily 1402.2808}}].

\bibitem{Hemming:2002kd}
S.~Hemming, E.~Keski-Vakkuri and P.~Kraus, \emph{{Strings in the extended BTZ
  space-time}},
  \href{https://doi.org/10.1088/1126-6708/2002/10/006}{\emph{JHEP} {\bfseries
  10} (2002) 006} [\href{https://arxiv.org/abs/hep-th/0208003}{{\ttfamily
  hep-th/0208003}}].

\end{thebibliography}\endgroup
\end{document}